\documentclass[sigconf]{acmart}

\usepackage{booktabs}    
\usepackage{tabularx}    
\usepackage{array}   
\usepackage{graphicx}
\usepackage{amsmath}
\usepackage{makecell}
\usepackage{hyperref}
\usepackage{multirow}
\AtBeginDocument{%
  }

\copyrightyear{2025}
\acmYear{2025}
\setcopyright{acmlicensed}\acmConference[UIST '25]{The 38th Annual ACM
Symposium on User Interface Software and Technology}{September 28-October
1, 2025}{Busan, Republic of Korea}
\acmBooktitle{The 38th Annual ACM Symposium on User Interface Software
and Technology (UIST '25), September 28-October 1, 2025, Busan, Republic of
Korea}
\acmDOI{10.1145/3746059.3747639}
\acmISBN{979-8-4007-2037-6/2025/09}

\begin{document}

\title[SCENIC]{SCENIC: A Location-based System to Foster Cognitive Development in Children During Car Rides}

\author{Liuqing Chen}\authornote{Corresponding author: Liuqing Chen,
\href{mailto:trovato@corporation.com}{chenlq@zju.edu.cn};
College of Computer Science and Technology, Zhejiang University}
\orcid{0000-0002-9049-0394}
\affiliation{%
  \institution{College of Computer Science and Technology, Zhejiang University}
  \city{Hangzhou}
  \state{Zhejiang}
  \country{China}
}
\email{chenlq@zju.edu.cn}

\author{Yaxuan Song}
\orcid{0009-0005-2664-4386}
\affiliation{%
  \institution{College of Computer Science and Technology, Zhejiang University}
  \city{Hangzhou}
  \state{Zhejiang}
  \country{China}
}
\email{songyx23@zju.edu.cn}

\author{Ke Lyu}
\orcid{0009-0000-6233-061X}
\affiliation{%
  \institution{College of Software Technology, Zhejiang University}
  \city{Ningbo}
  \state{Zhejiang}
  \country{China}
}
\email{lyuke@zju.edu.cn}

\author{Shuhong Xiao}
\orcid{0009-0007-0615-0607}
\affiliation{%
  \institution{College of Computer Science and Technology, Zhejiang University}
  \city{Hangzhou}
  \state{Zhejiang}
  \country{China}
}
\email{shuhongxiao@zju.edu.cn}

\author{Yilang Shen}
\orcid{0009-0002-7638-5712}
\affiliation{%
  \institution{College of Computer Science and Technology, Zhejiang University}
  \city{Hangzhou}
  \state{Zhejiang}
  \country{China}
}
\email{1195970116@qq.com}

\author{Lingyun Sun}
\orcid{0000-0002-5561-0493}
\affiliation{%
  \institution{International Design Institute, Zhejiang University}
  \city{Hangzhou}
  \state{Zhejiang}
  \country{China}
}
\email{sunly@zju.edu.cn}

\renewcommand{\shortauthors}{Chen et al.}
\acmArticleType{Research}
\acmCodeLink{https://github.com/borisveytsman/acmart}
\acmDataLink{htps://zenodo.org/link}
\acmContributions{BT and GKMT designed the study; LT, VB, and AP
  conducted the experiments, BR, HC, CP and JS analyzed the results,
  JPK developed analytical predictions, all authors participated in
  writing the manuscript.}

\begin{abstract}
Car-riding is common for children in modern life, and given the repetitive nature of daily commutes, they often feel bored, which in turn leads them to rely on electronic devices for entertainment. Meanwhile, the rich and rapidly changing scenery outside the car naturally attracts children’s curiosity, providing abundant resources for cognitive development. Our formative study reveals that parents' support during car rides is often fleeting, as accompanying adults may struggle to consistently provide effective guidance to nurture children's innate curiosity. Therefore, we propose SCENIC, an interactive system that guides children aged 6-11 to better perceive the external environment through location-based cognitive development strategies. Specifically, we built upon the experiential approaches used by parents, culminating in the formulation of six cognitive development strategies integrated into SCENIC. Additionally, considering the repetitive nature of car commutes, SCENIC incorporates features of dynamic POI selection and journey gallery generation to improve children's engagement. We evaluated the quality of SCENIC's generated content (N=21) and conducted an in-situ user evaluation involving seven families and ten children. Study findings suggest that SCENIC can enhance the car riding experience for children and help them better perceive the external environment through cognitive development strategies.
\end{abstract}

\ccsdesc[500]{Applied computing~Computer-assisted instruction}
\ccsdesc[300]{Human-centered computing~Ubiquitous and mobile computing systems and tools}
\ccsdesc[500]{Human-centered computing~User studies}

\keywords{Children; Car rides; User experience design; Education}

\maketitle

\section{Introduction}

In modern society, family rides with children have become a major headache for parents \citep{Wilfinger2011we}. The prolonged confinement in a limited car space leaves young children bored and fussy \citep{hoffman2013car, price2013travel, inbar2011make}, sometimes leading to emotional outbursts. These reactions not only cause tension and frustration within the family, but also pose a risk to driving safety \citep{eilitta2021children}. To address this issue, more and more parents rely on electronic devices to keep their children engaged during car rides. However, it has raised growing concerns over excessive screen time and its potential negative impact on children’s health and development \citep{gordon2015designing}.


In contrast to this growing view of car rides as a source of stress, previous studies regarded them as valuable opportunities for parent–child bonding \cite{hoffman2013car}. The confined space brings parents and children closer together, creating an uninterrupted environment for interaction and shared moments. Additionally, the scenes outside the car window offer rich visual stimulation \cite{kaplan1995restorative, faber2009children}, which spark children's curiosity and encourage verbal engagement with parents \cite{hiah2013engaging}. Theoretically, this dynamic visual environment is particularly beneficial for children aged 6-11 years, who are typically within Piaget's concrete operational stage of cognitive development. During this period, learning occurs most effectively through interactive, context-dependent experiences \citep{piaget1952origins, bruner1997culture, vygotsky1978mind, Piaget2005psychology, PiagetCognitiveDevelopment}. 

Nevertheless, this significant potential for cognitive development remains largely unrealized in practice, constrained by two evident limitations from the perspectives of both parents and children. First, the need to provide continuous, context-relevant communication and responses during car rides places a substantial multitasking burden on parents \cite{zhang2022storybuddy}, given their primary focus on driving safely. This often limits their ability to support children’s spontaneous questions or extend learning opportunities in real time. Second, the rapidly changing external environment during car rides can overwhelm children’s perceptual and cognitive capacities, causing them to miss meaningful stimuli or potential moments for reflection \cite{brunnberg2009games, tamminen2022parent, hoffman2013car}. Without appropriate scaffolding, these fleeting experiences fail to translate into sustained cognitive engagement.

Recognizing these challenges, this study explores how to design structured support that facilitates children’s cognitive development during car rides. As prior work in child-car interaction has not specifically focused on fostering broader cognitive development \cite{hoffman2013car, wu2020blokcar}, we first conducted a formative study involving eight families to understand the current state and dynamics of family rides. By summarizing parental practices for guiding children’s interactions with the external environment and identifying the challenges they face in this process, we introduced \textbf{SCENIC}: \underline{\textbf{S}}cenario-based role-playing, 
\underline{\textbf{C}}lassification, 
\underline{\textbf{E}}xpanded thinking, 
\underline{\textbf{N}}ormative self-regulation, 
\underline{\textbf{I}}nference, and 
\underline{\textbf{C}}onstrained choice. 
This framework consists of six cognitive prompting strategies that support learning during car rides by utilizing the learning potential embedded in surrounding environments. Leveraging the capabilities of generative AI to provide adaptive, real-time natural language interaction and access to a broad knowledge base \cite{nayeem2024kidlm, yan2024practical}, we integrated it into our approach to help sustain dialogue—mirroring the conversational style children are already familiar with during car rides. Specifically, we designed location-based storytelling grounded in passing scenery to foster engagement while reducing
parental burden. The system guides children to look out the window, listen to stories, initiate conversations, and receive answers to their questions through voice-based dialogue, thereby minimizing screen time.

In summary, this study presents the following contributions:
\begin{enumerate}
    \item \textbf{Formative Study Findings:} We identify how parents currently engage children with the external environment during car rides and the challenges they encounter.
    \item \textbf{Cognitive Prompting Strategy:} We propose a set of six strategies tailored for car rides to support children’s cognitive engagement using environmental cues.
    \item \textbf{System Implementation:} We design and develop a generative AI system that integrates cognitive prompts with location-based storytelling and journey gallery features.
    \item \textbf{System Evaluation:} We evaluated the quality of the generated-content in our system and carried out an in-situ user study, providing insights for designing location-based learning systems in transit for children.

\end{enumerate}

\begin{figure*}[htp]
    \centering
    \includegraphics[width=0.8\textwidth]{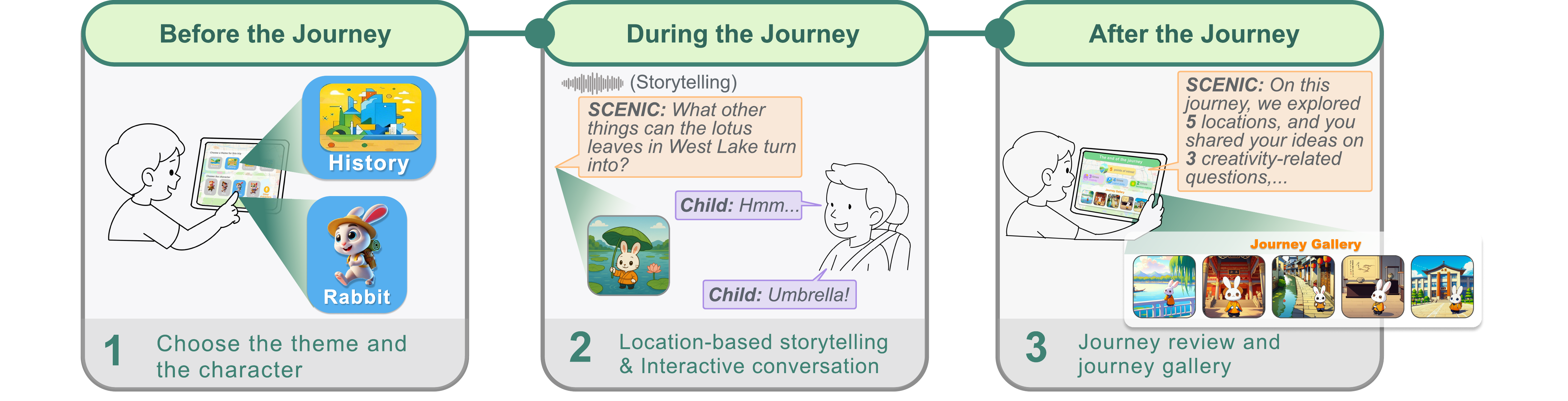}
    \caption{Overview of SCENIC: a system designed to support location-based cognitive development for children aged 6--11. Before the journey, children select a theme and character to personalize the experience. During the journey, interaction is primarily voice-based, involving location-based storytelling followed by an interactive conversation phase. In this phase, the system proactively delivers cognitive prompts based on the SCENIC strategies, while children can also freely ask their own questions and receive responses from the system. Image hints are provided if children encounter difficulty with system prompts. At the journey's end, SCENIC helps children review the locations interacted with and the number of cognitive prompts answered, and presents a journey gallery featuring the chosen character across these locations to reinforce memories.}

  \label{fig:scenic_overview}
    \vspace{-0.1in}
    \Description{Figure 1 is a three-part visual diagram showing the user journey in SCENIC, an interactive system for children. The first stage, “Before the Journey”, shows a child selecting a theme (e.g., History) and a character (e.g., a rabbit). The second stage, “During the Journey”, illustrates a location-based storytelling interaction, where the system asks a question about a scenery-related object and the child responds verbally. The final stage, “After the Journey”, depicts the child reviewing locations and responses and a journey gallery featuring the chosen character.}
\end{figure*}


\section{Related Works}
\subsection{Cognitive Development for Children Aged 6–11}
\label{RW_2.1}
To ground the design of systems for children, it is essential to consider their developmental characteristics. Our work targets children aged 6---11. Classically, this age range is often associated with Piaget's concrete operational stage, a period characterized by the emergence of logical thought about tangible objects, and a move away from egocentrism \citep{piaget1952origins}. Though Piaget theory has framed several HCI research \cite{vatavu2015touch}, we acknowledge that contemporary developmental science views cognitive growth as more continuous and variable than a strict stage model suggests, with many foundational abilities emerging earlier than Piaget proposed \citep{spelke1998nativism, meltzoff1988infant, gweon2010infants}. Acknowledging these limitations, our design does not adhere to a rigid stage theory. Instead, while using the 6---11 age range as a practical grouping, our design philosophy is broadly informed by theories of situational learning, which emphasize that knowledge acquisition is deeply intertwined with the context and activity in which it occurs \citep{lave1991situated, brown1989situated}.

Car rides present a promising context for situational learning, as the constantly changing scenery outside the windows offers rich, real-world content. However, much like other informal learning settings, in-car learning places great demands on the design of educational materials and the availability of adult support \cite{rogoff2016organization}. The car riding environment is both uncontrollable-cannot control the interaction span-and unpredictable-cannot anticipate what might appear next. These factors, compounded by the need for parents to prioritize driving, often result in inconsistent guidance. Consequently, in-car learning is more challenging than in other informal contexts—such as museums \cite{cronin2024factors, tenenbaum2010supporting}, in-home learning \cite{ho2024s}, and parks \cite{zhang2023observe}. In the field of Human-Computer Interaction (HCI), there remains a notable gap concerning the specific challenges and opportunities presented by in-car learning environments.

Due to the uncertainty of learning content in informal learning, there is no uniform standard for assessing children's cognitive development in this process. Nonetheless, several assessment indicators can be adapted to car riding environments. These include fact recall, which measures how well children remember passing points of interest (POIs), and productive memory processes, focusing on how children creatively build upon location-specific observations. In addition, evaluations based on within-journey conversations offer another valuable lens for understanding the impact of in-car informal learning.



\subsection{Interactions for Children in Car-riding Experiences}

Overall, the diversity of research on children’s in-car experience is relatively limited \cite{wu2020blokcar}, and most of them focus on helping children overcome boredom \cite{gordon2015designing, broy2011cooperative}. As a result, games have become the most common form of interaction. For instance, Gordon et al. \cite{gordon2015designing} proposed PANDA for interactive verbal games, while nICE \cite{broy2011cooperative} and Active Corners \cite{meschtscherjakov2016active} introduced collaborative games for multiple passengers, aiming to promote cooperation and communication. Although these entertainment systems have made a positive impact on reducing children’s boredom during car rides, they have often placed too much focus on the games themselves. This overemphasis can harm children’s eyesight, and user experiments have revealed an overlooked issue—namely, these systems fail to consider the journey experience itself \cite{broy2011cooperative}.

To address this issue, some studies have sought to redirect children’s attention outward by introducing location-based interactive innovations. For instance, Hoffman et al. \cite{hoffman2013car} implemented an augmented reality (AR) based collection game, Wu et al. \cite{wu2020blokcar} showcased block game results on car windows, and Brunnberg et al. \cite{brunnberg2009games} proposed an AR-based spatial perception shooting game. While these works capitalize on the location-specific attributes of the external environment-such as enhancing the sense of distance, location, and geographical scenes-and effectively enrich the child's travel experience, they remain largely short-term or entertainment-focused.

A notable commercial, Volkswagen's "Road Tales" in the Netherlands \footnote{\url{https://www.contagious.com/news-and-views/Volkswagen-The-Snelweg-Sprookjes}}, which is a mobile app that uses location data to trigger audiobooks for children corresponding to passing objects. However, its content is pre-defined and primarily entertainment-focused, lacking support for dynamic content generation or structured cognitive scaffolding. Building on these innovations, our approach addresses the limitations of static, entertainment-oriented content by leveraging Generative AI to create dynamic, personalized narratives with a structured framework of cognitive prompting strategies.

Overall, existing academic research on in-car experience design for children, while acknowledging educational objectives, primarily extends traditional entertainment methods that often come at the cost of prolonged screen time. Meanwhile, the potential of external environment as a resource for cognitive development remains largely underexplored in both research and real-world deployments. This study aims to transform the typically passive nature of car rides into an opportunity for children to actively observe, think about, and reflect their surroundings, thereby fostering cognitive development.


\subsection{Generative AI in Children Education}
As noted in Section \ref{RW_2.1}, the child-in-car-riding context is characterized by continuously changing scenery \cite{brunnberg2009games}, limited opportunities for physical interaction \cite{laurier2008driving}, and short interaction spans \cite{hoffman2013car}, with the added challenge of lacking instructor assistance \cite{hiah2013engaging}. These conditions make it essential for supporting tools to feature timely adaptability and engaging features to foster children's cognitive development in car rides. Motivated by these needs, we turned our attention to the advancements in generative AI-particularly large language models (LLMs) and text-to-image (T2I) models, and reviewed their existing applications in children’s education \cite{chen2024bidtrainer, nayeem2024kidlm, samala2025unveiling, chen2025mindscratch}.

Overall, the use of generative AI to promote children’s education and cognitive development has become increasingly prevalent in the past two years. One reason behind this trend is the extensive knowledge bases and natural language dialogue capabilities of these systems, which enable them to provide real-time, personalized feedback \cite{huber2024leveraging, ali2024picture, choi2024aacesstalk}. Building on these strengths, storytelling and Q\&A-based scaffolding have emerged as the two most widely designed strategies \cite{cai2025child, chen2025characterizing}, with researchers exploring their potential for fostering literacy \cite{dietz2024contextq, ye2024connection}, facilitating mathematical computation \cite{zhang2024mathemyths, xu2023mathkingdom, chen2024chatscratch}, and stimulating creativity \cite{chen2024chatscratch, han2023design, chen2024coremix}. In addition, some studies also explored how generative AI can enhance parent-child communication. For instance, StoryBuddy \cite{zhang2022storybuddy} highlights the tension parents face during picture book reading, as they aim for deeper engagement while also contending with the high cognitive load of multitasking. 

These applications offer valuable insights into how generative AI can be employed to support children's cognitive development during car rides. However, a cohesive framework encompassing both contextual cognitive development strategies and how generative AI could be used to enhance this process-remains underexplored. To address this gap, we first investigate the experiential approaches parents currently use to engage children with the passing environment, and then explore how generative AI can improve children's user experience in this process.

\section{Formative Study}
While existing literature provides some insights into the general state and dynamics of family car rides \cite{hoffman2013car, price2013travel, inbar2011make, tamminen2022parent, broy2011cooperative}, the specific role and utilization of the external environment as a resource for engagement and learning during these journeys remain relatively underexplored. Notably, while \citet{hoffman2013car} highlighted the need for parental effort to point out interesting locations, and another study observed children glancing outside when lacking internal companionship \citep{hiah2013engaging}, there is a scarcity of research investigating how families currently interact specifically around external scenery.


To better understand the current state and dynamics during family rides regarding the external environment, and how parents support their children’s inquisitiveness and cognitive development through interactions with the external environment, we conducted a formative study using semi-structured interviews with parents. Specifically, the primary goals of our formative study is to:
\begin{itemize}
    \item investigate the current state and dynamics of family interactions specifically related to the external environment during car rides;
    \item summarize parents' experiential approaches for leveraging the external environment for conversation or potential learning.
    \item identify the challenges that parents face in effectively guiding these interactions.
\end{itemize}

\subsection{Method}

\subsubsection{Participants.} Eight parents, each representing a unique family with at least one child aged 6-11 years, participated in this study. Demographic characteristics of each family are presented in Table~\ref{tab:family_info_formative}. The study focused on the experiences related to nine specific children within these families who met the target age range (6-10 years, M = 8.22, SD = 1.95). Among these nine children, five were female and four were male. Families were recruited via convenience sampling through the personal networks of two authors, and volunteered their participation. Each semi-structured interview, lasting approximately 40 minutes, was conducted with one parent per family via video conference.

\begin{table}[htbp]
\centering
\caption{Demographic characteristics of families participating in the formative study. Children aged 6--11 are highlighted in bold.}
\label{tab:family_info_formative}
\resizebox{\linewidth}{!}{%
\begin{tabular}{ccc}
\toprule
\textbf{Family ID} & \textbf{Participating Parent (Occupation)} & \textbf{Children Information} \\ \midrule
FF 1 & FP 1 (Female, Company Employee) & \underline{\textbf{10}} (female), 19 (female) \\ \midrule
FF 2 & FP 2 (Female, Designer) & \underline{\textbf{7}} (female) \\ \midrule
FF 3 & FP 3 (Female, Self-employed) & \underline{\textbf{10}} (male); 25 (female) \\ \midrule
FF 4 & FP 4 (Male, Engineer) & \underline{\textbf{7}} (female); 12 (male) \\ \midrule
FF 5 & FP 5 (Female, Sales) & \underline{\textbf{6}} (female); \underline{\textbf{10}} (male) \\ \midrule
FF 6 & FP 6 (Female, Manager) & \underline{\textbf{8}} (male) \\ \midrule
FF 7 & FP 7 (Female, Business) & \underline{\textbf{8}} (female); 14 (male) \\ \midrule
FF 8 & FP 8 (Male, Doctor) & \underline{\textbf{8}} (male); 12 (female) \\ 
\bottomrule
\end{tabular}%
}

\vspace{0.5em}
\raggedright
\small
\textbf{Notes:}
\begin{enumerate}
    \item FF represents a \textbf{F}ormative study \textbf{F}amily.
    \item FP represents the \textbf{F}ormative study Participating \textbf{P}arent.
\end{enumerate}
\end{table}



\subsubsection{Procedure.} The interviews were conducted by two of the authors. First, the researchers introduced the study's goals and obtained informed consent from the participants. After collecting basic information about the families, researchers conducted a semi-structured interview with the participating parent to explore the families' current in-car habits related to the external environment, their experiential approaches for leveraging the external environment, and the challenges they face in this process. We chose semi-structured interviews rather than in-situ observations during car rides, because our research topics are grounded in participants’ long-term experiences and sustained practices. The study was approved by the university’s ethics committee.

All interview sessions were audio-recorded for further analysis. After completing the eight interviews, the two researchers transcribed the recordings. Then, thematic analysis \cite{braun2006using} was employed to perform a qualitative analysis of these transcripts. Initially, the researchers independently coded the transcripts; these preliminary themes were later discussed and refined iteratively to reach a final consensus on the themes.

\subsection{Finding 1: Current Dynamics of Environmental Interaction During Family Rides}
Family car rides, occurring with frequencies ranging from daily (FF2, FF6) to weekly (FF5, FF8) for activities such as commuting, leisure outings, or visits, form the backdrop for our investigation. A key finding relates to children's natural curiosity and inquisitiveness with the external environment. Despite the potentially repetitive nature of routes, looking out the window was frequently cited as a common activity for children (6/8 families).

Parents reported that while children may often observe quietly, they also actively point out novel or interesting sights and proactively inquire about them (e.g., \textit{"My child often says: 'Mom, look, what’s that? What’s it used for?'"} -- FP6). Sometimes, parents also take the initiative to point out landmarks or provide information (FP1, FP3). This reveals a dynamic where both children and parents initiate interactions centered on the external environment.

However, sustaining this engagement proved challenging. Parents described children’s emotional states during rides as often fluctuating between periods of quiet observation and expressions of boredom, particularly during longer or more familiar routes (FP4, FP8). This boredom typically manifested as restlessness or complaints. While interaction related to the external environment holds the potential to mitigate this, difficulties in consistently maintaining effective engagement (due to factors summarized in Section~\ref{sec:formative_challenges}) often limit its impact. This contributes to the persistence of boredom, which, as many parents expressed concern, increasingly leads families to rely on electronic devices to manage children's restlessness, driven by the convenience of screen use and the growing dependence of children on technology in modern society.

\subsection{Finding 2: Parental Conversational Strategies Regarding the Environment}
\label{section_formative_experimential_approach}

Our interviews confirmed that conversation is the predominant form of interaction during family car rides. These conversations typically involved either routine daily dialogue (e.g., discussing school or family plans) or discussions specifically prompted by the external environment. As this study focused on the latter type, we analyzed parents' accounts to synthesize the conversational strategies they spontaneously employ when engaging with passing scenery.

 We categorized these conversational approaches according to the primary cognitive domains they appeared designed to stimulate or support. This analysis yielded three main categories of parental strategies: (1)~Fostering imagination, (2)~Enriching knowledge, and (3)~Cultivating choice expression. The subsequent paragraphs detail these categories with representative examples derived from the interviews.

\textbf{Fostering imagination.} This parental strategy involved using open-ended prompts, inspired by the external environment, to stimulate children's imagination. Parents encouraged forms of engagement such as making creative associations or engaging in perspective-taking (e.g., imagining the viewpoint or speech of an object). For example, interpreting shapes often prompted associations, as when one parent (FP4) asked about cloud formations: \textit{``What does that cloud look like?''} For an example of perspective-taking, when passed urban landscape sculptures in a square, FP8 had ask his child, \textit{``Look at the sculpture in the square, what do you guess they want to say?''} These techniques encouraged children to move beyond literal interpretation towards creative and imaginative thinking.

\textbf{Enriching knowledge.} This parental prompting strategy focuses on building children's foundational knowledge base by connecting observations to facts or concepts. We identified three common approaches within this category. Firstly, some parents (FP1, FP5, FP7) reported guiding their children to observe architectural styles of various buildings—such as banks, hospitals, and shops—to help them develop an understanding of the functions associated with different institutions. Secondly, parents attempted to leverage natural phenomena or basic mechanical principles encountered during the drive to explain scientific concepts(FP3, FP4, FP8). For instance, while one participant (FP3) described trying to explain car engine principles, she also highlighted a common challenge:\textit{``I had introduced the principles behind car engines to my son; however, I sometimes struggle to explain them in language that he can understand.``} Thirdly, seven out of the eight participants reported using roadside billboards, shop names, or street signs as opportunities to introduce new vocabulary, aiming to expand their children’s lexicons.

\textbf{Cultivating choice expression.} Offering choices related to the journey was another strategy parents employed to foster children's ability to articulate preferences. For instance, one parent (FP6) explained that during a family drive to a shopping mall, she discusses meal preferences with her child. Similarly, once on an outing to a botanical garden, another parent (FP7) asked her children where they wanted to go within the location. This type of prompt encourages children to weigh immediate options and make simple decisions regarding activities or specific places to visit.

\subsection{Finding 3: Challenges in Effective In-Car Environmental Conversation}
\label{sec:formative_challenges}

While parents widely recognize that conversations about the passing environment can alleviate children's boredom and potentially contribute to cognitive development, our findings indicate that sustaining such interactions effectively poses significant challenges for parents. Participants consistently described difficulties that primarily stem from two key areas: firstly, the limited attention due to multitasking inherent in the driving context, often compounded by other tasks; and secondly, the intrinsic difficulty of providing rich and varied conversations spontaneously in response to the dynamic environment.

\subsubsection{Limited Attention due to Multitasking.}
Parents consistently described how the primary task of driving significantly constrained their attention. Even non‑driving parents faced concurrent responsibilities—such as interacting with the driver and handling personal tasks or device use (occasionally work‑related)—which further divided their focus. This limited attentional capacity directly affected their ability to simultaneously monitor the environment for suitable sights, engage their child, and effectively nurture inquisitiveness. Moreover, parents noted that the cognitive effort required to spontaneously formulate relevant, developmentally appropriate questions and to process their child’s responses while managing other tasks added considerable pressure to maintaining consistent engagement.

\subsubsection{Challenges in Providing Rich and Varied Engagement.}
This challenge manifests in two primary ways: Firstly, limited parental knowledge about the diverse external scenery often constrained the depth and breadth of interaction. While children frequently expressed inquisitiveness, parents acknowledged struggling to identify or provide information about many specific locations or features encountered (FP3, FP6, FP8). Consequently, especially on familiar or repetitive routes, the conversational prompts parents could offer tended to become repetitive, potentially failing to sustain children's interest and possibly even exacerbating boredom.

Secondly, the nature of the interaction was often shaped by parental biases and personal interests, rather than necessarily aligning with pedagogical goals or the child's curiosity. Lacking broad knowledge or specific strategies, parents naturally drew upon their own expertise or interests when formulating questions or comments. This could lead to a narrow focus, influenced by factors such as professional background (e.g., FP2, a designer, mentioned had discussed traditional Chinese patterns inspired by a passing unique architecture), which might not consistently match the child's own interests. This reliance on familiar personal topics limited the diversity of cognitive domains addressed and the types of questions asked, hindering the potential for multi-faceted cognitive stimulation.

\subsection{Design Goals}
Informed by the findings from our formative study---which highlighted the challenges parents face in sustaining engagement and the untapped potential of children's environmental curiosity---we established the following three design goals to guide the development of our system:
\begin{itemize}
    \item[\textbf{G1:}] \textbf{Foster Sustained Engagement.} The system should provide an engaging and enjoyable experience that can capture and maintain children's interest throughout the car ride.
    \item[\textbf{G2:}] \textbf{Facilitate Knowledge Acquisition about the Environment.} The system should support children in learning about the environment by providing prompts that are cognitively beneficial and developmentally appropriate, and directly relevant to the locations encountered during the riding.
    \item[\textbf{G3:}] \textbf{Encourage Active Interaction with the External Environment.} Beyond just presenting information, the system should actively prompt children to observe, think about, and reflect the external scenery, thereby deepening their interaction with and awareness of the world.
\end{itemize}

\section{The Design of SCENIC}
This section details the design of SCENIC, a location-based system employing generative AI to foster children's cognitive development during car rides. A primary goal of SCENIC is to direct children's attention towards the outward scenery, aiming to channel their natural curiosity into cognitively engaging interactions related to their surroundings. The system is designed for children aged 6---11, an age range informed by cognitive development theories \cite{piaget2013play} and insights from our formative study. SCENIC operates on touchscreen tablet devices and utilizes both touch and voice input to provide an accessible and intuitive interaction experience within the car environment. All user study interfaces were presented in their original Chinese and translated into English for this manuscript.

The design process, detailed in the following subsections, was primarily informed by our formative study investigating current parent-child interactions and challenges during car rides.

First, based on the identified opportunities for cognitive scaffolding, we developed the SCENIC cognitive prompting strategies (Section~\ref{sec_SCENIC_strategies}). These strategies form the pedagogical core of the system, designed to foster specific cognitive skills by encouraging active reflection about the environment (\textbf{G2}). 

However, presenting these prompts in isolation could feel disjointed. Therefore, to create a more immersive and engaging interaction experience (\textbf{G1}, \textbf{G3}) and to provide context for the cognitive prompts (\textbf{G2}), we designed a location-based storytelling framework (Section~\ref{sec_storytelling}). This framework weaves the cognitive prompts into narrative episodes, with each episode anchored to a specific real-world location.

To select suitable locations to serve as these interaction anchors while ensuring variety and novelty across journeys, we developed a dynamic Point of Interest (POI) selection mechanism (Section~\ref{sec:POI_selection}) (\textbf{G1}). In SCENIC, POIs are specific, map-identified locations (such as a museum or park) that trigger the story episodes and the cognitive prompts. 

Finally, the experience is further enhanced by supporting features like the journey gallery, designed to reinforce memories and prolong engagement (\textbf{G1}).

\subsection{SCENIC: Cognitive Prompting Strategies}
\label{sec_SCENIC_strategies}




Our formative study revealed that conversation is the predominant interaction approach parents use to engage their children with the external environment. This dialogue is bidirectional, with parents proactively asking questions and children posing their own inquisitive queries. Consequently, our system adopts this established and natural conversational interaction patterns. 

Building on the three identified categories of parental conversational strategies—fostering imagination, enriching knowledge, and cultivating choice expression—SCENIC is designed to support children’s development in creativity, logical reasoning, and decision-making. However, our analysis showed that parents' spontaneous questions, while valuable, often lacked cognitive depth and variety. When examined through the lens of Bloom's Taxonomy \citep{forehand2010bloom}, most parent-generated prompts focused on lower-order thinking skills such as Remembering (e.g. \textit{``What is the name of this building?"} and Understanding (e.g. \textit{``Why do people visit this tower?"}).

\begin{figure*}[htp]
    \centering
    \includegraphics[width=0.9\textwidth]{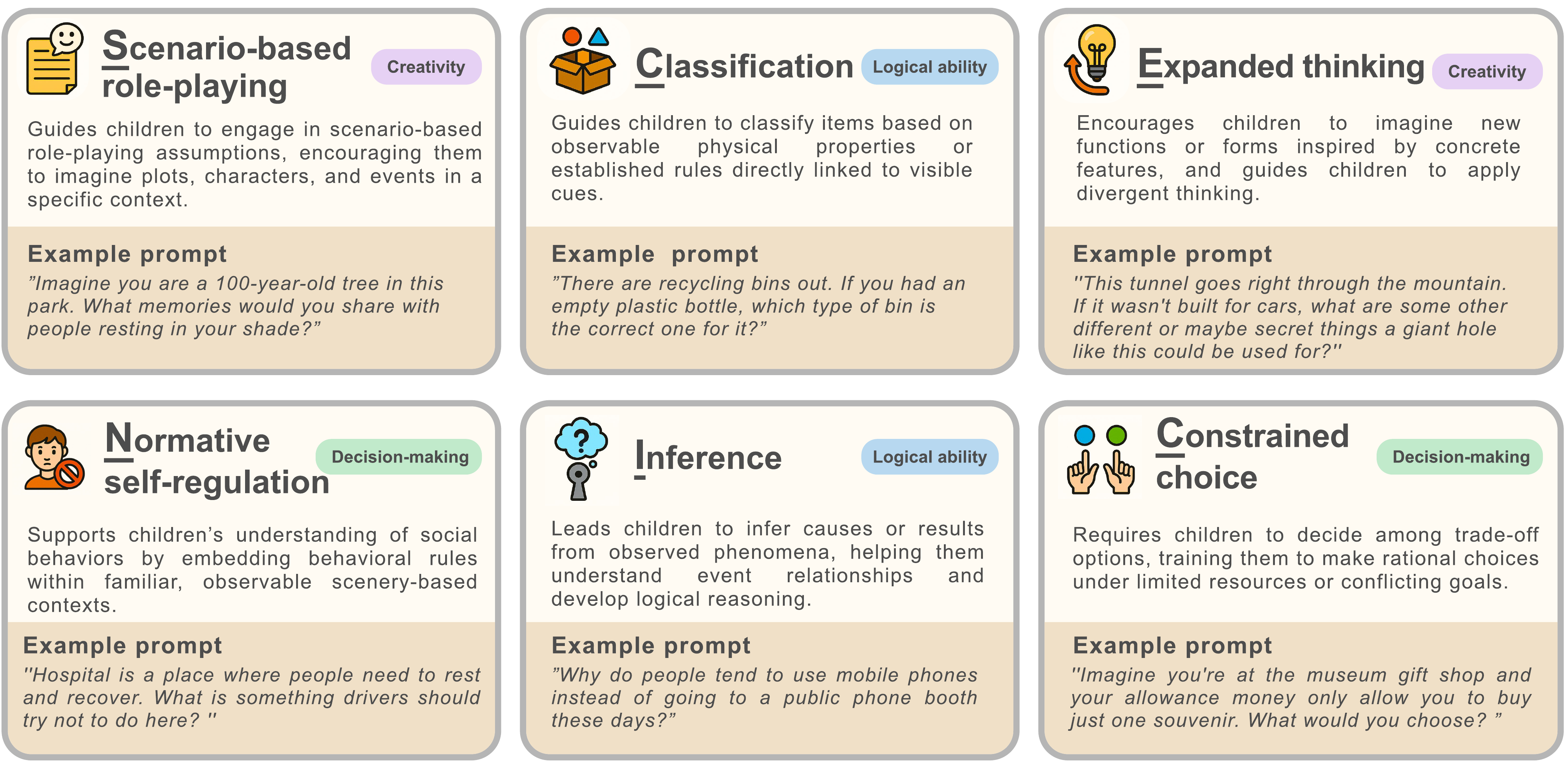}
    \caption{Overview of the SCENIC cognitive prompting strategies designed to foster cognitive development during car rides. The diagram details the six strategies: Scenario-based role-playing, Classification, Expanded thinking, Normative self-regulation, Inference, and Constrained choice. Each strategy is presented with its description, targeted developmental goal (creativity, logical ability, or decision-making), and an example prompt.}
    \label{fig:SCENIC_strategies}
     \vspace{-0.1in}
     \Description{Figure 2 shows six cognitive prompting strategies used in the SCENIC system to support children’s cognitive development during car rides. Each strategy is presented as a labeled card containing the strategy name, a brief description, its developmental goal (creativity, logical ability, or decision-making), and an example prompt. The six strategies include: Scenario-based role-playing, Classification, Expanded thinking, Normative self-regulation, Inference, and Constrained choice.}
\end{figure*}

To address this limitation and provide a more structured scaffolding of cognitive skills, we designed SCENIC strategies: a set of six location-based cognitive prompting strategies, with a particular emphasis on fostering higher-order thinking skills such as Analyzing, Evaluating, and Creating \textbf{(G2)}. Each strategy (Figure~\ref{fig:SCENIC_strategies}) is grounded in established pedagogical theories and explicitly anchored to concrete, observable objects or situations encountered during the ride. This design serves a dual purpose: supporting children's situated learning through concrete thinking during the operational stage, and constraining prompt generation to avoid overly abstract or generic outputs common in LLMs \cite{zhang2024mathemyths, chen2024chatscratch}.

The following paragraphs provide detailed explanations for each of the six strategies.

\textbf{Scenario-based role-playing:} As one strategy to foster creativity, this approach leverages the external scenery to guide children to engage in scenario-based role-playing assumptions, encouraging them to imagine plots, characters, and events in a specific context. For instance, when passing a park, SCENIC might ask: \textit{“Imagine you are a 100-year-old tree in this park. What memories would you share with people resting in your shade?”} By prompting children to step outside their own viewpoint and embody concrete elements of the scene (e.g., observed objects), this strategy targets creativity while aiming to cultivate empathy and connection to the environment. 

\textbf{Classification:} Aiming to foster logical reasoning, this strategy guides children to classify items based on observable physical properties or established rules directly linked to visible cues. For example, upon observing different colored collection bins placed curbside near a crossroads, SCENIC might pose a multiple-choice question: \textit{``See the recycling bins out. If you had an empty plastic bottle, which type of bin is the correct one for it? Is it (A) The Recycling bin, (B) The regular Trash bin, or (C) The Compost bin?''} This prompts the child to apply classification rules to a concrete object based on visible environmental features (the bins), reinforcing logical classification.

\textbf{Expanded thinking:} As another strategy targeting creativity, this approach guides children to apply divergent thinking, focusing on specific, observable objects or structures within the scenery. It encourages them to imagine new functions or forms inspired by concrete features. For example, when approaching or driving through a tunnel, SCENIC might ask: \textit{``This tunnel goes right through the mountain. If it wasn't built for cars, what are some other different or maybe secret things a giant hole like this could be used for?''} Such prompts encourage imaginative associations based on observed physical characteristics, moving beyond conventional uses.

\textbf{Normative self-regulation:} This strategy supports children’s understanding of social behaviors by embedding behavioral rules within familiar, observable scenery-based contexts. It utilizes prompts encouraging reflection on expected actions relevant to specific, encountered scenarios or locations. For example, when the system detects the car is passing near a hospital (a concrete place), SCENIC might ask: \textit{``A hospital is a place where people need to rest. What is something drivers should try not to do here? (A) Drive carefully, (B) Make lots of loud noise, or (C) Follow traffic signals?''} By situating regulations within a tangible real-world setting, such prompts help children internalize norms through relatable, concrete examples.

\textbf{Inference:} This strategy fosters logical reasoning by guiding children to understand cause-and-effect relationships related to the purpose or function of specific structures or places. It encourages inferring reasons based on observable context or general knowledge about why things are built or designed in a certain way. For instance, when the system identifies the car is approaching or crossing a large bridge, SCENIC might ask: \textit{“Look at this bridge we are crossing! Why do you think people needed to build such a long bridge right here?”}. Engaging with prompts that require explaining the 'why' based on the likely function of a structure in relation to its location helps children practice inference skills about purpose and necessity.

\textbf{Constrained choice:} This strategy helps children practice decision-making using realistic scenarios often tied to specific, encountered locations or objects. It presents choices involving concrete items or activities, requiring children to weigh tangible factors. For example, prompted by passing a museum, SCENIC might ask: \textit{``Imagine you're at the museum gift shop with enough allowance for just one souvenir. Would you choose (A) a model dinosaur skeleton kit, (B) a colorful book about planets, or (C) a soft plush toy mascot? Which one would you pick, and why?"} This supports evaluation skills and understanding trade-offs related to desirable, concrete options.

\subsection{Location-based Storytelling}
\label{sec_storytelling} 




To reduce the abruptness of posing questions directly at each location, we designed a location-based storytelling framework that introduces narrative context prior to the interactive prompts. This narrative approach aims to enhance contextual immersion \textbf{(G1, G3)} and provide a natural pretext for the interactive prompts \textbf{(G2)}.

\begin{figure*}[htp]
    \centering
    \includegraphics[width=0.9\textwidth]{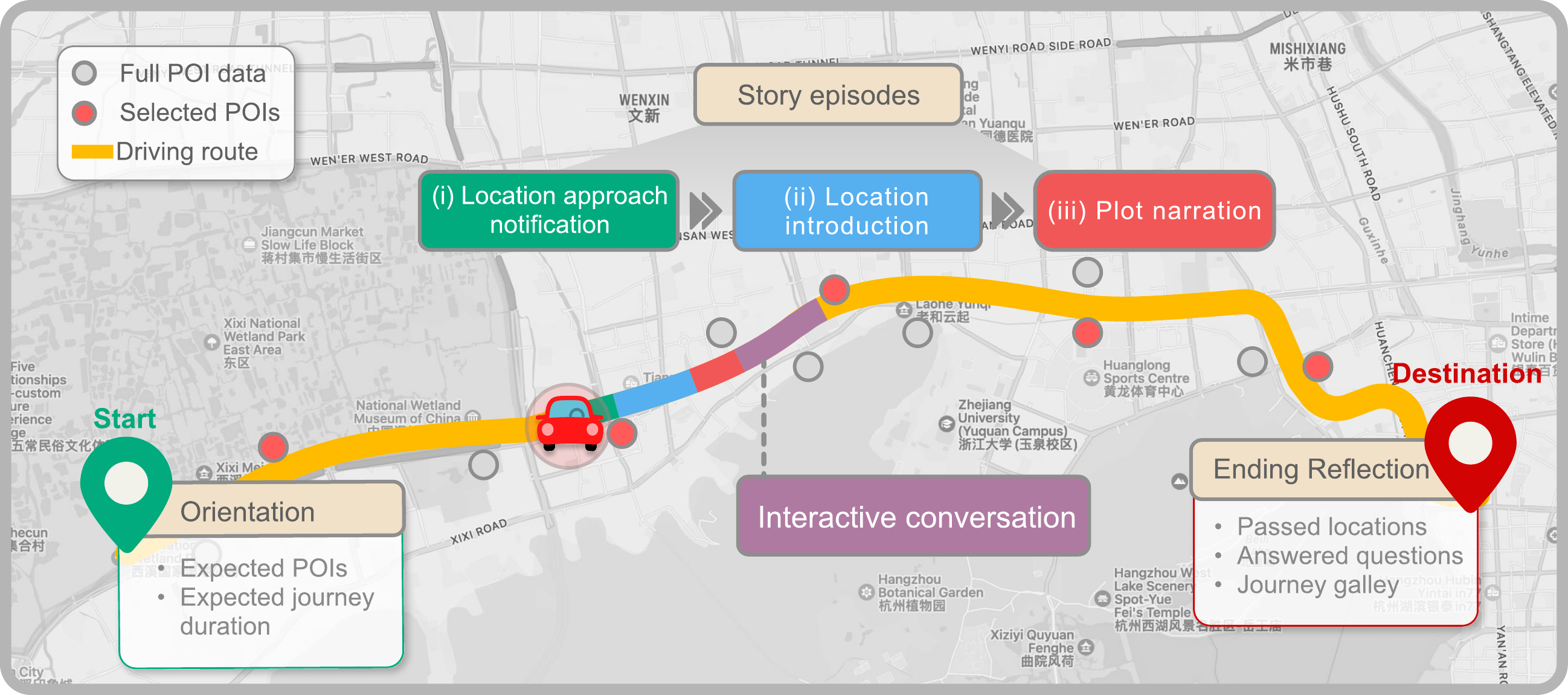}
    \caption{Overview of the location-based storytelling process, comprising POI selection (based on POI type, spacing, and route distance), the location-based storytelling (including the location approach notification, location introduction, and plot narrative).}
    \label{fig:POI-selection}
    \vspace{-0.1in}
     \Description{Figure 3 shows an overview of the location-based storytelling process in the SCENIC system. The visual depicts a driving route with selected POIs marked along the path. The process begins with an orientation segment introducing expected POIs and journey duration. As the vehicle approaches each POI, three sequential stages are shown: (i) location approach notification, (ii) location introduction, and (iii) plot narration. Throughout the journey, interactive conversations occur. The route ends with an “Ending Reflection” phase, where children revisit passed locations, answered questions, and journey highlights. A legend explains the icons for full POI data, selected POIs, and driving route.}
\end{figure*}

As illustrated in Figure~\ref{fig:POI-selection}, our location-based storytelling process comprises several key design components: (a) \textbf{POI selection}, which considers factors such as POI type, spacing, and the whole route distance; (b) \textbf{Story episodes} at each location, which includes: (i) the location approach notification designed to guide children to observe the scenery outside the car window, (ii) the location introduction tailored to the child's selected character, and (iii) the location narrative that builds the story around the child's chosen character at each location. Besides, to enhance personalization and engagement and introduce variety to repetitive car rides---especially during daily commutes, SCENIC allows pre-journey selection of a preferred story theme and main character. Furthermore, the system dynamically adapts story background descriptions based on real-time weather conditions, aiming to increase congruence between the narrative and the child's actual visual experience through the car window. The following subsubsections provide a detailed description of these features.

\label{section_design_story}

\subsubsection{Story Structure and Narrative Design}
Our storytelling process employs a structured narrative format designed for the in-car context, comprising three main components: an Orientation, multiple Episodes (Episode 1, Episode 2,\dots, Episode n), and an Ending Reflection. 

The Orientation phase initiates the experience. The system introduces the chosen protagonist and confirms the narrative theme with the child. It also provides context for the journey by previewing the approximate number of key locations to be featured and indicating the estimated travel duration (data sourced from Amap).

Each subsequent Episode is anchored to a specific location identified along the route. The narrative within an episode typically unfolds in three parts: (1)~a notification that the vehicle is approaching the location, ideally timed to direct children's attention outwards (e.g., triggered approximately 100 meters prior; (2)~an introduction of relevant background information about that place, sourced from reliable external data (Wikipedia\footnote{\url{https://en.wikipedia.org/wiki/Main_Page}}) to mitigate LLM hallucination; and (3)~the plot narration involving the protagonist’s actions or experiences at that location. Following the narration, the system initiated the cognitive prompting, and dynamically adjust the number of prompts presented to flexibly adapt to varying distances between locations or driving conditions like traffic congestion or traffic light stops.

Finally, the Ending Reflection phase consolidates the journey, summarizing metrics such as the number of locations the child interacted with via the system and the number of cognitive prompts answered. Accompanying this summary is the journey image gallery. This gallery presents a sequence of visuals, each depicting one of the interacted-with locations, incorporating the child’s chosen character within the scene. This visual recap enables children to revisit the journey's highlights, reinforcing their memories and connection to the experienced locations.

To improve the internal quality of the story, we have implemented three measures. First, the themes and characters chosen by the children will serve as clues to connect the potentially disjointed episodes, enhancing the consistency of the story. Second, smooth transitions are crucial to maintaining children's immersion, especially when significant environmental changes occur between locations (e.g. "the [protagonist] enters the campus"). To improve the language appropriateness for children that is easy to follow in a moving vehicle, we adopted a few-shot strategy by referencing five children's picture books (designed for ages 6-11) from an open-source website\footnote{\url{https://www.oxfordowl.co.uk/}}. This approach encourages simpler, clearer sentences, the use of metaphors and similes, and avoids overly complex concepts. For further details on the specific prompt design, please refer to Supplementary material A.


\subsubsection{POI selection}
\label{sec:POI_selection}
To ensure that the locations selected to anchor the narrative and interactive prompts are both suitable and dynamically varied (\textbf{G1}), our system employs a multi-stage POI selection algorithm. This algorithm begins with established Geographic Information System (GIS) methods for identifying potential candidates and then applies a series of SCENIC-specific filtering criteria related to POI type and spatial distribution.

The core process involves several key steps. First, using the Amap API and the planned route, potential POIs are identified ensuring they are likely to be visible from the roadside by children in the car. Second, this initial set is filtered based on POI type labels provided by the API, removing categories deemed unsuitable for children's cognitive engagement or those that are excessively commercialized. The remaining POIs are then subject to further criteria regarding spacing and density. Through iterative design, we established the following POI selection criteria. (1) No POIs are selected within 1 km from the origin and destination, reserving time for journey orientation and reflection. (2) A minimum spacing of 800 meters between any two POIs is maintained to allow adequate time for narrative storytelling and interactive conversations. (3) Selected POI types must be mutually exclusive to enrich the diversity of information encountered by children during their travel. (4) The maximum number of POIs is constrained based on route distance to minimize cognitive overload for children. Specifically, for journeys shorter than 10 km, up to 4 POIs are selected; for journeys between 10 and 15 km, up to 5 POIs; and for journeys between 15 and 20 km, up to 6 POIs are included.

\subsection{Implementation}
Overall, SCENIC comprises a front-end web page built in React and a back-end server developed in Python. To avoid literacy barriers, we utilized the Tencent Cloud Text-to-Speech API\footnote{\href{https://cloud.tencent.com/product/tts}{https://cloud.tencent.com/product/tts}} for voice agent to broadcast the voice prompts and the react-speech-recognition\footnote{\href{https://github.com/JamesBrill/react-speech-recognition}{https://github.com/JamesBrill/react-speech-recognition}} for transcribing speech. On the front-end, we leverage the AMap Locate API\footnote{\href{https://lbs.amap.com/product/search/}{https://lbs.amap.com/product/search/}} for location detection, presenting real-time context-related stories and initiating POI-specific real-time interactions when entering the detection range of a POI. On the back-end, the AMap Search API\footnote{\href{https://lbs.amap.com/product/locate/}{https://lbs.amap.com/product/locate/}} is utilized to automatically select POIs through polygon search and filtering algorithms. Relevant information, integrated with prompt engineering, is then fed to LLMs via the Deepseek-V3 API\footnote{\href{https://platform.deepseek.com/}{https://platform.deepseek.com/}} to generate location-based storytelling content, SCENIC questions, and real-time interaction feedback. We selected Deepseek-V3 based on its strong performance in handling Chinese--essential for our target context---and its API latency. During the real-time interaction phase, LLMs process inputs from children and provide appropriate responses by combining corresponding story content and questions. Additionally, DALL-E\footnote{\href{https://openai.com/index/dall-e-3/}{https://openai.com/index/dall-e-3/}} is employed to generate images depicting POI-related questions, aiding children in immersing themselves in the context and understanding the questions. We obtain the estimated driving time through AMap Path API\footnote{\href{https://lbs.amap.com/product/path}{https://lbs.amap.com/product/path}}, and use Gemini 2.0 Flash Experimental\footnote{\href{https://ai.google.dev/gemini-api/docs/image-generation}{https://ai.google.dev/gemini-api/docs/image-generation}} to generate the journey gallery.

\subsection{User Flow}

Overall, SCENIC supports independent child use with minimal parental involvement. To illustrate the SCENIC system workflow, we use the experience of Lucky, a seven-year-old boy, during his 16 km trip to a history museum with his parents on a Saturday.

\begin{figure*}[htp]
    \centering
    \includegraphics[width=0.8\textwidth]{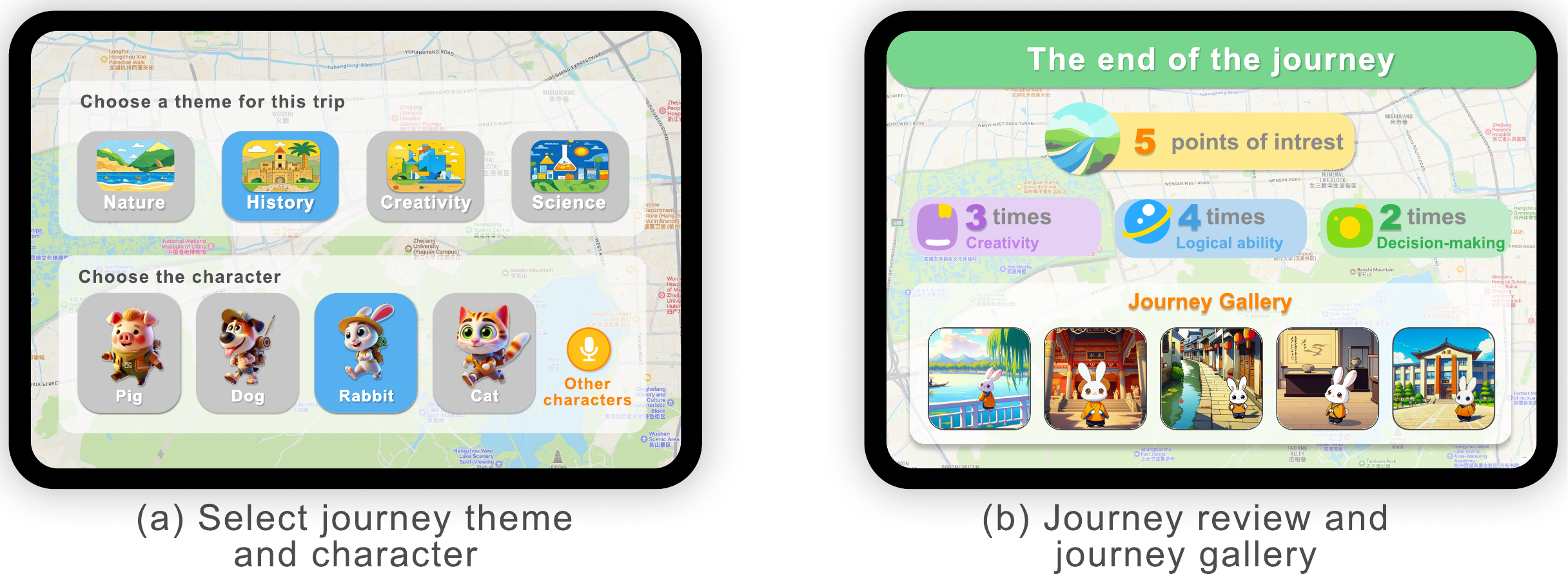}
    \caption{SCENIC system user interfaces for pre-journey setup and post-journey reflection. (a) Before the ride, children personalize their experience by selecting a journey theme and character. (b) After the ride, the Ending Reflection screen summarizes the journey, displaying interacted-with locations, the number of cognitive prompts answered per category (creativity, logical ability, decision-making), and a journey gallery featuring visuals of these locations incorporating the chosen character.}
  \label{fig:scenic_ui_flow}
    \label{fig:UI}
    \vspace{-0.1in}
     \Description{Figure 4 presents two main user interface screens of the SCENIC system. The left image (a) shows the pre-journey setup screen, where children choose a journey theme from four options—Nature, History, Creativity, and Science—and select a character such as Pig, Dog, Rabbit, or Cat to accompany them. The right image (b) displays the post-journey reflection screen titled “The end of the journey,” summarizing the child’s experience. It highlights the number of cognitive prompts answered per category (3 for creativity, 4 for logical ability, 2 for decision-making), and shows a journey gallery with visual snapshots of locations visited, featuring the chosen character (Rabbit) interacting with each scene.}
\end{figure*}

Before the journey begins, the initial screen (Figure~\ref{fig:UI} (a)) allows children to select their preferred journey theme and character. The system offers four themes—nature, history, creativity, and science—which shape the narrative style. The chosen character subsequently serves as the protagonist in the generated story. For Lucky, he selected ``History`` and "Rabbit" for the subsequent journey.

During the journey, there is no need for children to focus on the screen, and the user interface (show in Supplementary material B) only displays the route with marked POIs and the estimated remaining travel time to help reduce time-related anxiety among children (mentioned by the parents in the formative study). It is necessary to emphasize that the primary interaction throughout the journey is interactive conversation, and the children can use the voice wake-up function to respond. Using the voice wake-up function allows for hands-free interaction, enabling children to engage without needing to physically interact with the screen, which enhances safety and convenience during the journey. Children are only expected to glance at the screen when visual cues (i.e. image illustrations) are needed to help them understand the question.

As the journey begins, the system first plays the Orientation segment of the story, previewing the upcoming POIs: \textit{``Bunny Duoduo's historical exploration journey has begun! Today, she will travel back in time through our city. Over approximately [Estimate] minutes, she will uncover secrets at five locations: Su Causeway, the Deshou Palace, a historic cultural block, the Ceramics Museum, and a university, before finally arriving at the destination: the Natural History Museum.}

Subsequently, when the vehicle is approximately 100 meters from the first location, the system initiates a pre-notification: \textit{``Look ahead! We are approaching the famous Su Causeway''}, capturing the child’s attention towards the location outside. A brief background introduction, is then provided: \textit{``This long path across the lake was built by a famous poet named Su Shi almost a thousand years ago! Bunny Duoduo imagines people walking here in ancient clothes, enjoying the view...''} Following this, the system narrates the in-situ story plot: \textit{``Bunny Duoduo hops along the Su Causeway, feeling the gentle breeze from the lake, and thinks about the poet who built this beautiful path…''}

After the story episode concludes, the system transitions into the interactive conversation session. For example, a cognitive prompt based on expanded thinking might be: \textit{``What other things can the lotus leaves in West Lake turn into?"} If children experience difficulties in answering, such as long pauses or proactively asking for help (\textit{“I have no idea.”}), the system would offer hints through image illustrations. For instance, the system might generate an image of the bunny using lotus leaves as an umbrella, helping children form ideas for their response.

Moreover, the span between POIs is flexibly controlled by the number of interactive conversation items. Sometimes, the system answers questions raised by children, while at other times, it encourages children to think about the external environment through questions, until the vehicle is approaching the next location. Throughout the journey, interactions are primarily based on episodes that loop, utilizing voice-based exchanges.

At the end of the journey, the system presents a Journey Review and Journey Gallery (see Figure~\ref{fig:UI}(b)). 

Collectively, the design components in SCENIC are intended to support sustained engagement, even across familiar and repetitive commutes. The dynamic POI selection algorithm introduces variety in the locations encountered, while the inherent variability of the generative storytelling and the multi-faceted nature of the SCENIC prompting strategies ensure that even revisited locations can be experienced in novel ways.

\section{Generated-content Quality Evaluation}
\subsection{Method}
\subsubsection{Data Collection}
For the evaluation of generated-content, we curated three routes representative of common journeys for children within a city in Southern China. We designate these routes using the prefix GCR (\textbf{G}enerated-\textbf{C}ontent \textbf{R}oute). Each route originates from a different residential community and leads to a distinct destination: GCR1 refers to a journey to a shopping mall (approximately 17~km), GCR2 to a public park (approximately 9~km), and GCR3 to an art museum (approximately 14~km). Applying the POI selection algorithm in our system, there are six, four, and five POIs for GCR1, GCR2, and GCR3, respectively.

To evaluate the cognitive prompting strategies, three POIs were randomly selected from each GCR. We compared three question sets: (1)~SCENIC-generated, (2)~Parent-authored, and (3)~Baseline LLM (Deepseek-V3 without SCENIC strategies). We collected the Parent-authored question set by re-contacting participants from our formative study and conducting online interviews. Parents were asked: \textit{``When passing this location, what questions would you ask your child?"} During data collection, we observed that individual parents often found it challenging to spontaneously generate six distinct questions for a single location. Therefore, leveraging the insights and examples gathered across all parent interviews, the final Parent-authored set was compiled to represent a diverse collection of questions reflecting how parents might typically engage their child (aged 6---11) regarding the nine locations identified across the three GCRs. For the Baseline LLM set, Deepseek-V3 was prompted with the details of these nine locations using a generic instruction (see Supplementary material C). SCENIC then generated its corresponding prompts for the same nine locations.

For the evaluation of the location-based storytelling compassed two aspects: the overall quality of the narratives and the child-appropriateness of the story language. For the overall quality of the narratives, we prompted SCENIC to generate one complete story sequence for each combination of the three routes (GCR1, GCR2, GCR3) based four themes. Each story sequence consisted of an Orientation segment followed by interconnected Episode segments. The final Reflection segment, which is tied to children’s interaction data, was excluded from the this evaluation. In total, this process produced 12 complete story sequences (3 routes × 4 themes = 12).


Storytelling evaluation encompassed narrative quality and language appropriateness. For narrative quality, SCENIC generated 12 complete story sequences (excluding the final reflection which requires the children interaction data) covering all combinations of the 3 GCRs and 4 predefined themes. For language appropriateness, we compared paired story episodes generated for 12 representative POIs (4 from each GCR). Using a single theme ('Nature Exploration', selected as representative), we generated episodes with SCENIC's few-shot adaptation (A) and the baseline LLM without adaptation (B). This A/B design aimed to isolate the linguistic impact of our adaptation, assuming minimal influence from the specific theme choice on core language metrics.

\subsubsection{Data Analysis}
We recruited two groups of evaluators for the evaluation of the generated content. The first group, responsible for rating cognitive prompt quality and overall story quality, consisted of 18 parents recruited via an online survey platform (with at least one child aged 6-11) and 3 education researchers identified by the authors. Each evaluator in this group was asked to rate a balanced subset of materials, which included three sets of generated questions employing three strategies (SCENIC, Parent-authored, Baseline LLM) based on three locations, as well as four SCENIC-generated stories. Ratings were provided using five-point Likert scales.


For the evaluation of the location-based cognitive prompting strategies, we used a multi-faceted approach. First, we assessed the prompts using four key metrics: location relevance, cognitive significance, developmental appropriateness, and question diversity. A detailed explanation of each metric is provided in Supplementary material E. Evaluators were given clear definitions outlining the precise scope and intent of each metric prior to rating. The first three metrics (Location Relevance, Cognitive Significance, Developmental Appropriateness) were rated for each individual question within the three comparison sets. In contrast, Question Diversity was assessed once per group for the entire set of questions associated with each specific location. For the statistical analysis, we first tested the data for normality. Depending on the outcome, an ANOVA or the non-parametric Kruskal-Wallis H test was utilized based on data distribution to detect statistically significant differences in the metric ratings between the three groups. Post-hoc tests with Bonferroni correction were used for pairwise comparisons where appropriate.

Second, to further provide a theory-grounded classification of cognitive depth, we conducted a question-level analysis using Bloom’s Taxonomy \cite{forehand2010bloom}. This framework classifies cognitive questions into six hierarchical levels, from lower-level (Remember, Understand, Apply) to higher-level (Analyze, Evaluate, Create), with higher levels indicating deeper cognitive engagement \cite{raz2023role}. Two education researchers independently coded each question to one of these six levels. The initial inter-rater reliability was substantial, with a Cohen's Kappa of 0.78 \cite{landis1977measurement}. All remaining disagreements were subsequently resolved through discussion to reach a final consensus for analysis. A chi-square test was then used to compare the distribution of questions across these cognitive levels between the three groups.

For the evaluation of location-based story quality, we assessed internal narrative coherence and appeal using three primary metrics. Plot consistency measured the logical flow and the absence of contradictions between consecutive story episodes linked to different locations along the route. Theme relevance evaluated the extent to which the overall narrative arc aligned with the designated story theme provided to evaluators in the questionnaire. Engagement reflected the story’s perceived ability to capture and sustain children’s interest. 

It should be noted that the language appropriateness evaluation was conducted with a distinct group of 11 parents, separate from those who evaluated the overall story quality. This separation was implemented to mitigate potential bias, ensuring that judgments about language suitability were not influenced by prior exposure during the story quality assessment. In addition to the quantitative ratings, we included open-ended questions to collect qualitative feedback explaining the rationale behind the parents' choices.

All evaluators participating in the evaluation tasks were provided with monetary compensation commensurate with the time they spent to complete their assigned assessments. The study protocol was approved by the university’s ethics committee.

\subsection{Results}
\subsubsection{SCENIC Cognitive Prompting Strategies Evaluation}
Table~\ref{tab:Q&A_evaluation_results} summarizes the evaluation results for the cognitive prompting questions. Overall, significant differences between the SCENIC, Parent-authored, and LLM-only groups were found for location relevance ($p<0.001^{**}$), cognitive significance ($p<0.001^{**}$), and question diversity ($p=0.03^{*}$). Post-hoc analysis using Bonferroni correction revealed that SCENIC significantly outperformed both the Parent-authored and LLM-only groups on location relevance (all $p < 0.01^{**}$) and cognitive significance (all $p < 0.01^{**}$). For question diversity, SCENIC also scored significantly higher than the Parent-authored group ($p = 0.01^{**}$) and the LLM-only group ($p = 0.014^{**}$). Additionally, the SCENIC prompts received the highest average rating (M = 4.59) among the three methods, suggesting a positive trend in perceived suitability for the target age group compared to the other methods.

\begin{table}[ht]
\centering
\caption{Evaluation results for contextual relevance, cognitive significance, developmental appropriateness, and question diversity.}
\label{tab:Q&A_evaluation_results}
\footnotesize
\setlength{\tabcolsep}{3pt}
\begin{tabularx}{0.45\textwidth}{lcccc}
\toprule
\multirow{2}{*}{\textbf{Group}} & \multicolumn{4}{c}{\textbf{Mean (SD)}} \\
\cmidrule(l){2-5}
& \makecell{\textbf{Location}\\ \textbf{Relevance}}
& \makecell{\textbf{Cognitive}\\ \textbf{Significance}}
& \makecell{\textbf{Developmental}\\ \textbf{Appropriateness}}
& \makecell{\textbf{Question}\\ \textbf{Diversity}} \\
\midrule
\textbf{SCENIC}          & 4.61 (0.72) & 4.54 (0.75) & 4.59 (0.82) & 4.73 (0.57) \\
\textbf{Parent-authored} & 4.31 (0.92) & 4.24 (1.06) & 4.40 (0.98) & 4.19 (1.08) \\
\textbf{LLM-only}        & 4.38 (0.82) & 4.30 (0.90) & 4.35 (0.91) & 4.38 (0.85) \\
\midrule
\textbf{P-value}         & 0.001\textsuperscript{**} & 0.001\textsuperscript{**} & 0.1194 & 0.03\textsuperscript{*} \\
\bottomrule
\end{tabularx}

\vspace{0.5em}
\raggedright
\footnotesize
\textbf{Note:} ** denotes $p < 0.01$, * denotes $p < 0.05$.

\end{table}

\begin{table*}[ht]
\centering
\caption{Percentage distribution of cognitive question types across conditions based on Bloom's Taxonomy. Lower-order = Remember, Understand, Apply; Higher-order = Analyze, Evaluate, Create.}
\label{tab:bloom_percentage_summary}
\begin{tabularx}{\textwidth}{l c c c c c c | c c}
\toprule
\textbf{Condition} 
& \textbf{Remember} 
& \textbf{Understand} 
& \textbf{Apply} 
& \textbf{Analyze} 
& \textbf{Evaluate} 
& \textbf{Create} 
& \textbf{Lower-order} 
& \textbf{Higher-order} \\
\midrule
\textbf{SCENIC}           
& 11.1\% & 5.6\% & 5.6\% & 24.1\% & 31.5\% & 22.2\%  
& 22.3\% & 77.8\% \\

\textbf{Parent-authored}  
& 53.7\% & 16.7\% & 9.3\%  & 7.4\% & 9.3\% & 3.7\%   
& 79.7\% & 20.4\% \\

\textbf{LLM-only}         
& 25.9\% & 29.6\% & 18.5\% & 5.6\% & 7.4\% & 13.0\%  
& 74.0\% & 26.0\% \\
\bottomrule
\end{tabularx}
\end{table*}

A chi-square test of independence was performed to examine the relationship between the question source (SCENIC, Parent-authored, LLM-only) and the cognitive level classification. The analysis revealed a statistically significant association between the source of the question and its cognitive level, $\chi^2$(2, N = 162) = 44.64, $p < .001$.

Given the significant overall result, pairwise comparisons were conducted using chi-square tests with a Bonferroni-adjusted significance level (adjusted $\alpha = .0167$). The results indicated that the proportion of higher-order questions generated by SCENIC was significantly greater than that of the Parent-authored group, $\chi^2$(1, N = 108) = 35.61, $p < .001$, and also significantly greater than that of the LLM-only group, $\chi^2$(1, N = 108) = 29.08, $p < .001$. No significant difference was found between the Parent-authored and LLM-only groups, $\chi^2$(1, N = 108) = 0.47, $p = 0.494$.

As shown in Table~\ref{tab:bloom_percentage_summary}, these statistical differences are reflected in the descriptive data: 77.8\% of SCENIC's questions were classified as higher-order, in contrast to 26.0\% for the LLM-only group and just 20.4\% for the Parent-authored group. These findings suggest that SCENIC was more effective at eliciting cognitively demanding questions that extend beyond factual recall or surface-level understanding.

\subsubsection{Story Quality}
SCENIC-generated stories were evaluated positively for their internal quality. On a five-point Likert scale, mean ratings were consistently high for theme relevance (M = 4.71, SD = 0.55), plot consistency (M = 4.62, SD = 0.62), and engagement (M = 4.56, SD = 0.68). This suggests that evaluators found the narratives generated by SCENIC to be generally coherent, aligned with their designated themes, and capable of capturing children's interest.



Regarding the evaluation of language appropriateness for children, results from the paired comparisons (SCENIC with few-shot adaptation vs. baseline LLM without adaptation) showed a clear preference for the adapted versions: 62.12\% of evaluator choices favored the content generated using few-shot learning with children's picture books. Qualitative feedback highlighted the reasons for this preference, with evaluators frequently noting that the adapted versions employed more ``straightforward language'', ``short sentences'', and ``emotionally appealing expressions''.

This difference is illustrated by an example generated for passing an AI-driven home design company. The baseline model's content included relatively technical terms such as ``3D modeling'', ``intelligent algorithms'', and ``rendering previews''. In contrast, the few-shot adapted version used more concrete and imaginative terms like ``3D models'' and evoked relatable concepts such as designing a ``fantasy park'', demonstrably aligning better with young children's cognitive and linguistic levels. A detailed illustration could be found in the Supplementary material D.

\section{User Evaluation}
To assess the effectiveness of SCENIC promoting children's interactions of the external environment, we conducted a short-term evaluation focusing on children's engagement with SCENIC and the feedback from children and parents. Our study was conducted in-situ rather than in a laboratory setting, in order to capture more realistic design and usability issues. Overall, we aimed to address three research questions:
\begin{itemize}
    \item \textbf{RQ1:} How does SCENIC contribute to an engaging car-riding experience for children?
    \item \textbf{RQ2:} How does SCENIC facilitate children's perception and knowledge of the external environment?
    \item \textbf{RQ3:} How does SCENIC influence children’s interactions with the external environment?
\end{itemize}

\subsection{Method}
We recruited 7 families involving 10 children (C1–C10; 4 females, 6 males; age range 6–10, M=7.5, SD=1.26) through online advertisements. Demographic characteristics of each family are presented in Table~\ref{tab:user_study_family_info}.

We opted against a within-subjects comparative design (e.g. driving the same route with and without SCENIC), primarily to mitigate child's potential fatigue or disengagement associated with lengthy or repetitive driving sessions. Thus, each family was asked to select a driving route for the study that they had previously traveled together with their child. All families received compensation and institutional souvenirs based on their participation time (the total duration of the three experimental sessions), approximately 100 RMB per hour, with each participant averaging 1.5 hours. The study was approved by our institution.

\begin{table}[htbp]
    \centering
    \caption{Demographic information of families participating in the user study.}
    \label{tab:user_study_family_info}
    \resizebox{\linewidth}{!}{
    \begin{tabular}{ccccc}
        \toprule
        \textbf{Family ID} & \textbf{Child ID} & \textbf{Age} & \textbf{Gender} & \textbf{Participated Parent(s) ID} \\ \midrule
        UF 1  & C1  & 6  & male   & M1, F1 \\ \midrule
        UF 2  & C2  & 6  & male   & M2, F2 \\ \midrule
        \multirow{2}{*}{UF 3} 
            & C3 & 8  & female & \multirow{2}{*}{F3} \\ 
            & C4 & 8  & female &  \\ \midrule
        UF 4  & C5  & 7  & female & M4, F4 \\ \midrule
        \multirow{2}{*}{UF 5}
            & C6 & 8  & male   & \multirow{2}{*}{M5} \\ 
            & C7 & 10 & female &  \\ \midrule
        \multirow{2}{*}{UF 6}
            & C8 & 6  & male   & \multirow{2}{*}{M6, F6} \\ 
            & C9 & 8  & male   &  \\ \midrule
        UF 7  & C10 & 8  & male   & M7, F7 \\ 
        \bottomrule
    \end{tabular}
    }

    \vspace{0.5em}
    \raggedright
    \textbf{Notes:}
    \begin{enumerate}
        \item UF represents a \textbf{U}ser Study \textbf{F}amily.
        \item Participating parents are labeled as \textbf{M} (Mother) or \textbf{F} (Father) followed by the family number (e.g., M1 = Mother from UF 1).
    \end{enumerate}

\end{table}

\subsection{Procedure}
\label{sec:user_procedure}
\begin{figure}[htp]
    \centering
    \includegraphics[width=0.9\linewidth]{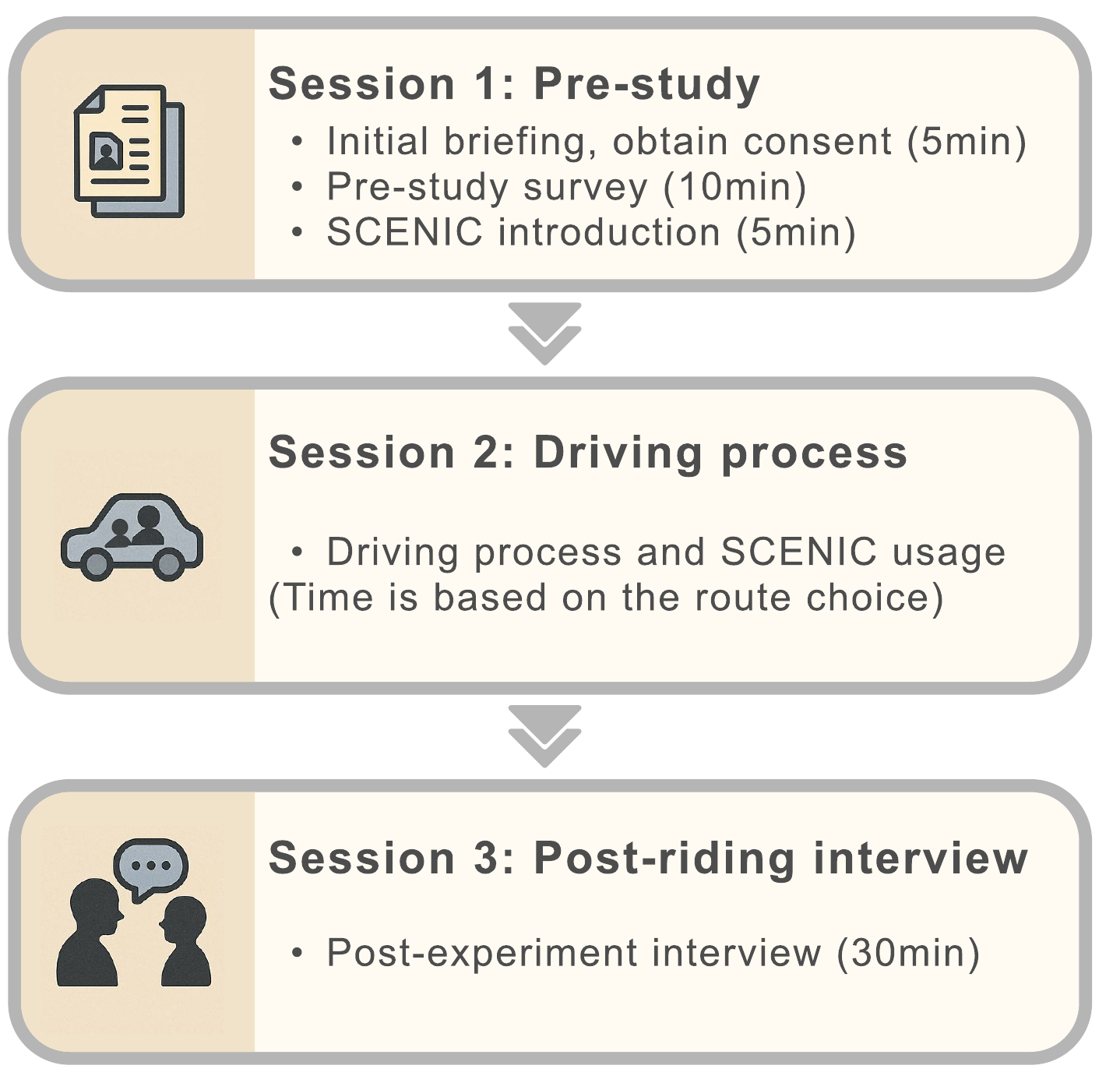}
    \caption{Experimental procedure of the in-situ user study.}
    \label{fig:enter-label}
     \vspace{-0.1in}
     \Description{Figure 5 illustrates the experimental procedure of the in-situ user study, consisting of three sequential sessions. Session 1: Pre-study includes an initial briefing, informed consent, a 10-minute pre-study survey, and a 5-minute introduction to the SCENIC system. Session 2: Driving process involves participants using SCENIC during a real-world car ride, with session duration dependent on the selected driving route. Session 3: Post-riding interview is a 30-minute post-experiment interview to reflect on the experience.}
\end{figure}

Each family participated in three sessions (see Figure~\ref{fig:enter-label}): the first served as a pre-study survey, the second focused on SCENIC usage during driving, and the third comprised post-experimental assessments and semi-structured interviews.

\subsubsection{Session 1: Pre-study}
In the initial session, we provided a brief introduction to the study objectives, addressed any questions, and obtained informed consent. Next, we confirmed the experimental route with each family and re-confirmed the familiarity. Parents were asked about the typical locations they might point out or discuss with their children. We recorded the number of locations they mentioned. Besides, children were interviewed briefly about their existing knowledge of the route, including any specific locations or landmarks they could recall and any notable features they remembered noticing along the way.

Finally, before commencing the drive, we introduced the SCENIC system to the child and guided them through selecting their preferred journey theme and character to personalize the upcoming experience.

\subsubsection{Session 2: Car-riding Process}

\begin{figure}[htp]
    \centering
    \includegraphics[width=0.6\linewidth]{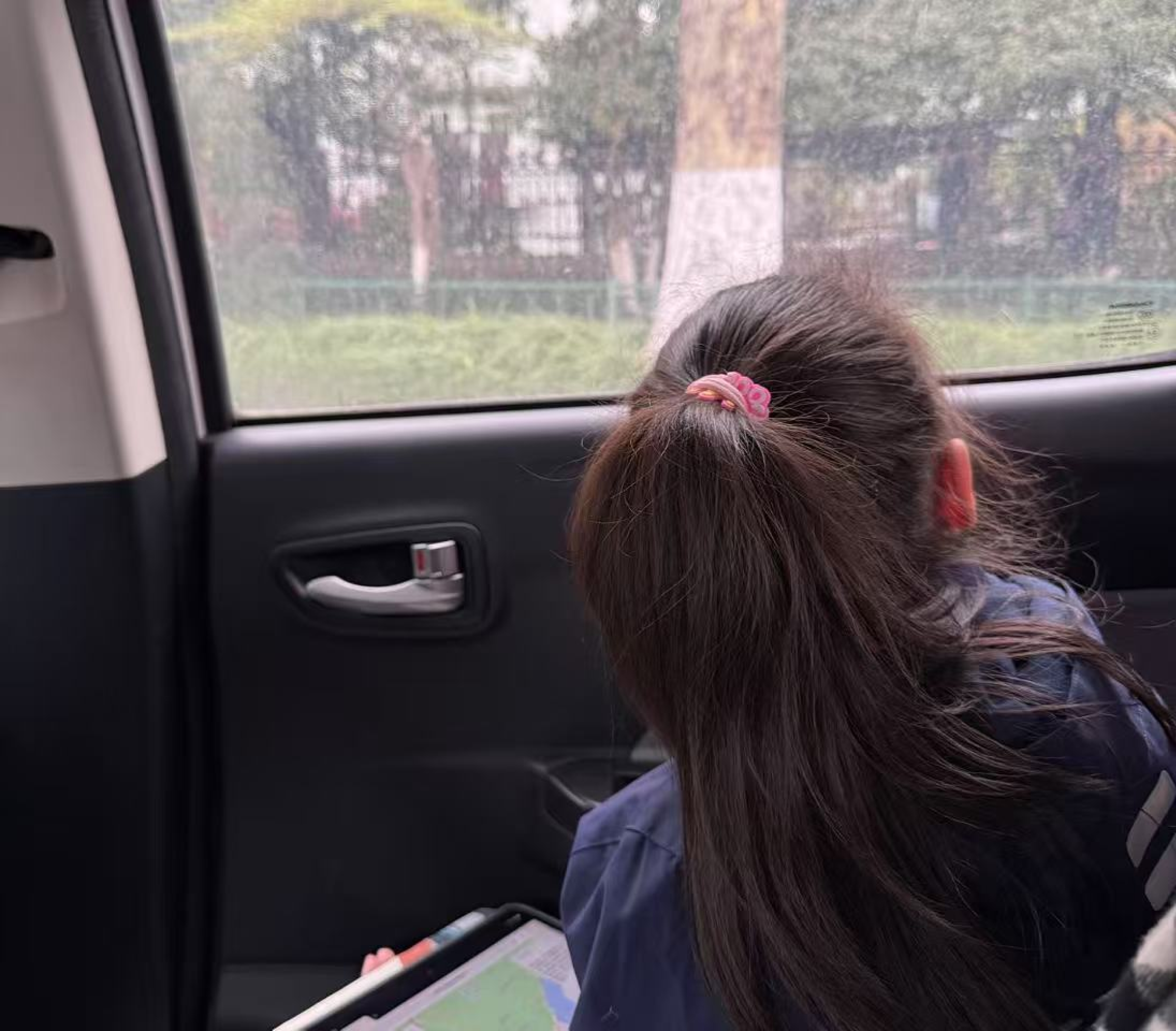}
    \caption{Photo of a participated child looking outside the car window.}
    \label{fig_photo}
    \vspace{-0.1in}
    \Description{Figure 6 shows a photo taken during the in-situ user study. A child participant is seated in the backseat of a car, looking outside the window. }
\end{figure}

During the user study drive, the child participant was seated securely in the rear seat, while one parent served as the driver. One primary researcher sat alongside the child in the rear seat to facilitate and observe children's interaction with the SCENIC system, which was running on a tablet. To ensure safety and manage the interaction effectively, this researcher generally held the tablet or placed it securely nearby, utilizing mainly the system's voice interface. When SCENIC presented specific visual aids (e.g., story illustrations) that required direct viewing, the researcher would position the tablet clearly for the child to see briefly, before returning it to a secure position. At the same time, this researcher served as the primary observer, recording the child's notable behaviors, verbal responses, and engagement patterns using detailed field notes. We did not restrict parental intervention during the drive to maintain a naturalistic setting. A photo of a participated child looking outside the car window is shown in Figure~\ref{fig_photo}.

\subsubsection{Session 3: Post-riding Study}
Upon reaching the destination, we first conducted semi-structured interviews and administered the Giggle Gauge questionnaire (detailed in~\ref{sub_data_collection}) to assess the children’s experiences and recollections of the journey. Subsequently, we interviewed the parents to gather their attitudes toward their children’s interactions with our system and to solicit suggestions for further system improvements.


\subsection{Data Collection and Analysis}
\label{sub_data_collection}
In this section, we describe the data collection and evaluation methods in our user study. Notably, we collected data from each child independently if the family have multiple children. 

\subsubsection{Parental Locations Nomination and Child Route Recall}

As mentioned in Section~\ref{sec:user_procedure}, two key baseline measures were collected during the pre-study session regarding the chosen familiar route. For parent-nominated locations, we asked: \textit{``Please list all the locations you might point out to your child on this route."} For the child's long-term recall, we asked: \textit{``Your parent said you’ve traveled this route before. What places or features do you remember along the route?"}

\subsubsection{Children's Giggle Gauge Instrument.} We adopt Giggle Gauge \cite{dietz2020giggle} instrument for evaluating children's experience with SCENIC, as it is a commonly adopted metric for investigating children's engagement with technology \cite{zhang2022storydrawer, dietz2023visual}. It assess children's attitudinal engagement in terms of seven perceptions including aesthetics, challenge, control, endurability, feedback, interest, and novelty on a 4-point Likert scale, thereby enabling children in our target age group to express their feelings more easily. Full definitions and justifications of each item are provided in the Supplementary material E.

\subsubsection{Interaction Logs and Usage Video Recordings.}
In this study, we collected two types of recordings. One type captured the story content and interactions (text and images) between children and SCENIC, while the other recorded the children's physical behaviors during the car-driving session. From the interaction logs, we analyzed whether children's responses were based on the location-based storytelling, the generated image for inspiration, or their own experience, and we also tallied the questions, distinguishing those asked by the children from those posed by SCENIC. For the behavioral recordings, we particularly focused on the duration and frequency of children's gazes out of the car window, especially when the system audio stated, \textit{"We are approaching the [POI]."} Reflexive thematic analysis \cite{braun2021one} was applied to both data sources.

\subsubsection{Post-Drive Semi-structured Interviews.} We conducted separate semi-structured interviews with the children and parents. 

Interviews with the children aimed to capture their direct perspectives on using SCENIC and its impact on their journey experience and memory recall. We asked questions designed to elicit both general impressions and specific memories of the trip, while also probing the role of different system components.

For short-term recall, we asked: \textit{``Could you tell me about the places we passed on our drive?"} Follow-up questions included: \textit{``What details do you remember about [specific recalled place]?"} and \textit{``What helped you remember those details—was it the story, the questions, the pictures, or something else?"} Finally, to assess their perception of SCENIC’s influence, we asked: \textit{``How was using SCENIC today different from your usual car rides?"} This approach allowed us to gather rich, qualitative insights into each child's subjective experience, accommodating the variability in routes and locations across different families.

Interviews with parents focused on their evaluation of SCENIC, their attitudes towards it, and suggestions for further improvement. Example questions included: \textit{"Did you feel that your child was focused while using SCENIC?"}, \textit{"Did you believe that your child appeared happy during the usage?"}, and \textit{"Would you be willing to support your child in continuing to use this system?"}

\subsection{Results}

\subsubsection{RQ1: How does SCENIC Contribute to an Engaging Car‑Riding Experience for Children?}
\mbox{}\\
Firstly, we examined how SCENIC engaged children with the passing POIs. Results from the Giggle Gauge questionnaire indicated that children showed an overall “moderate” engagement level (M = 3.51, SD = 0.30), according to the established interpretation range (3.0–3.6) \cite{dietz2020giggle}. Compared to the neutral benchmark score of 3.0, this represents a large effect size (Cohen’s d = 1.74).

Breaking down the scores further, we found that the endurability (M = 3.90, SD = 0.32), interest (M = 3.8, SD = 0.422), and novelty (M = 3.90, SD = 0.32) ratings were all above 3.6, suggesting “high” engagement levels \cite{dietz2020giggle}. This indicates that children enjoyed using SCENIC and were likely to use the system again in the future. However, the feedback score (M = 2.80, SD = 0.63)\footnote{The “Feedback” item (“The app let me know when I did something”) specifically reflects momentary interaction responses, rather than indicating overall system responsiveness.} was relatively low, which may be attributed to occasional delays in voice response during system usage.

Children’s engagement with SCENIC was further supported by their enthusiastic responses during the semi-structured interviews. For instance, when asked about their favorite features, more than half of the participants (C1-C5, C8-C10) specifically mentioned the final journey photo collection function. One child (C10) exclaimed, \textit{"I want to save these photos because they are precious memories of my journey."} Another child (C1) described it as “so heart-warming,” especially since his favorite rabbit appeared in one of the photos. These quotes illustrate the “wow effect” experienced by many children during interactions \cite{giusti2018framework}.

It is important to note that the visuals shown in the journey gallery are directly linked to the locations the child interacted with during the car ride. A more comprehensive view of how different design features contributed to children's engagement across the entire route is provided in Figure~\ref{fig:sankey} in Section~\ref{section_RQ2_result}.

\subsubsection{RQ2: How does SCENIC Facilitate Children's Perception and Knowledge of the External Environment?}
\label{section_RQ2_result}
\mbox{}\\
\textbf{Comparison of POI selection: SCENIC vs. Parent nomination.} 
We first compared the number of POIs automatically selected by the SCENIC system against the number of POIs nominated by parents prior to the drive. Data were collected across eight route observations (as Family 6 undertook two distinct routes, each involving one child using the system while the other observed). Table~\ref{tab:route_data} presents the descriptive statistics.

A Mann-Whitney U test was conducted to compare the number of POIs selected by SCENIC versus those nominated by parents pre-drive. The results revealed a statistically significant difference (SCENIC: M = 5.25, SD = 1.28; Parent-nominated: M = 2.13, SD = 1.13; p < 0.01). This indicates that SCENIC identified significantly more POIs along the routes compared to the parents' pre-drive selections.

Qualitative analysis suggests potential reasons for this difference. Parents often focused on nominating prominent landmarks or destinations, typically places they had previously visited with their children and held established familiarity. For instance, the parent in Family 6, being highly familiar with the destination park, nominated multiple points within that area. Furthermore, in several cases (Families 1, 2, and 5), parents nominated only the final destination as the single anticipated POI for conversation along the entire route.

In contrast, SCENIC identified a significantly larger and more diverse set of POIs. Critically, this included locally relevant locations previously unknown to parents, despite being situated along familiar routes. This capability offers distinct advantages: firstly, it facilitates the discovery of new, potentially interesting places for future family outings, as highlighted by one parent's comment on finding a previously unknown library for children: \textit{``I had no idea there was a children’s library nearby; it had provided us with a promising new destination for future outings.''} Secondly, by identifying more granular POIs throughout the journey, SCENIC creates opportunities for more frequent and continuous interaction, enriching the travel experience beyond a singular focus on the destination. 

\begin{table}[htp]
\centering
\caption{Routes in our user experiment and the number of POIs selected by parents' pre-drive and those recommended by SCENIC.}
\label{tab:route_data}
\resizebox{0.45\textwidth}{!}{%
\begin{tabular}{lccc}
\toprule
\textbf{Route} & \textbf{Parents' nomination} & \textbf{SCENIC} & \textbf{Driving time} \\ 
\midrule
Route 1 (UF1)    & 1                      & 5               & 32min               \\
Route 2 (UF2)    & 1                      & 4               & 20min               \\
Route 3 (UF3)    & 3                      & 8               & 47min                \\
Route 4 (UF4)    & 2                      & 5               & 36min                \\
Route 5 (UF5)    & 1                      & 4               & 34min                \\
Route 6 (UF6)    & 3                      & 5               & 28min                \\
Route 7 (UF6)    & 4                      & 5               & 36min                \\
Route 8 (UF7)    & 2                      & 6               & 39min                \\ 
\bottomrule
\end{tabular}
}
\end{table}

\begin{figure}[htp]
    \centering
    \includegraphics[width=0.95\linewidth]{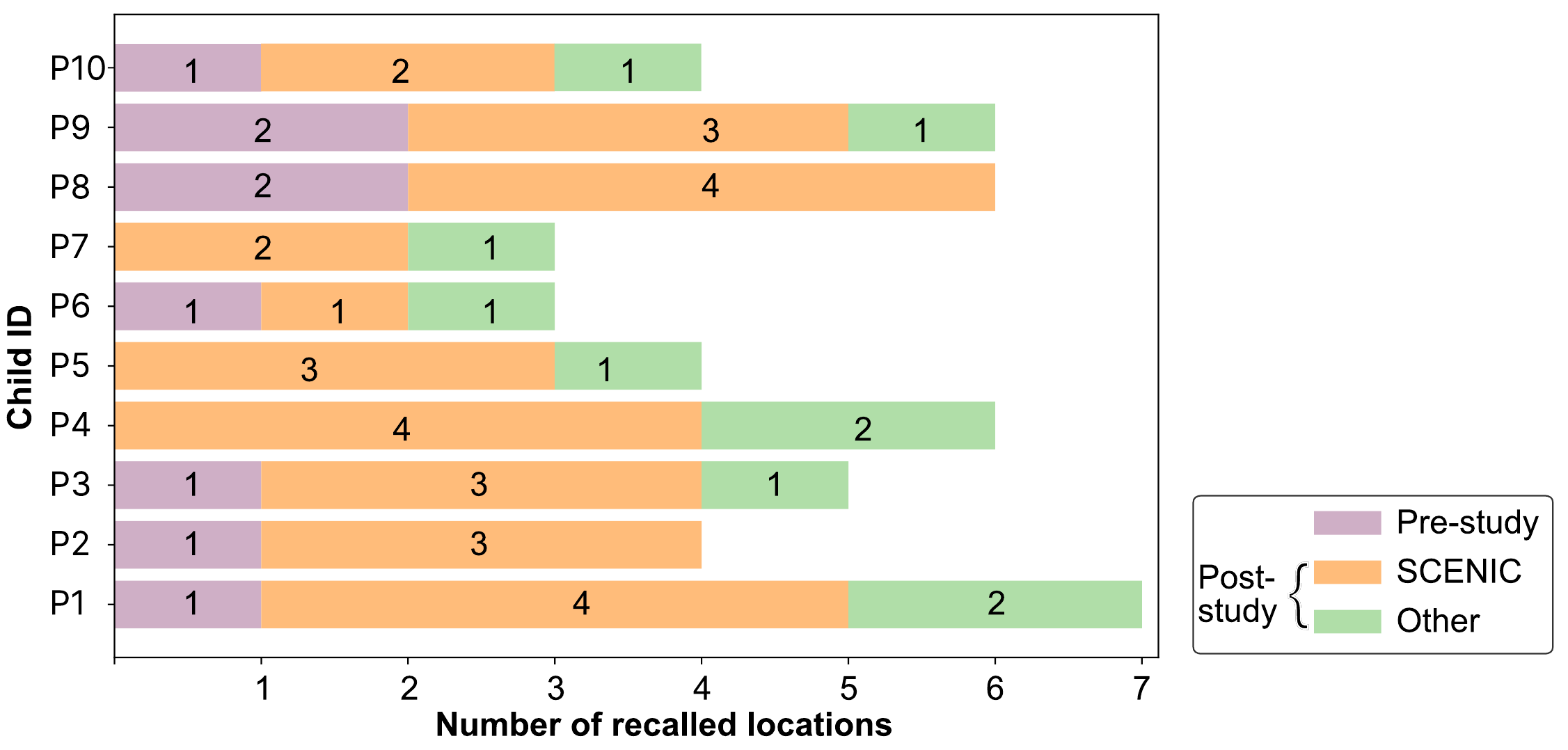}
    \caption{Comparison of the number of locations investigated in the pre-study (purple color coded), and those recalled after the experiment (orange color coded represents SCENIC-recommended locations, green represents other locations reported by the children).}
    \label{fig:number_POI_child}
    \vspace{-0.1in}
    \Description{Figure 7 presents a horizontal stacked bar chart comparing the number of locations associated with each child participant before and after the experiment. Each bar represents one child (P1 to P10), showing the total count of recalled locations. The purple segment indicates locations nominated during the pre-study by parents. The orange segment shows locations recalled post-study that were recommended by the SCENIC system. The green segment reflects other locations spontaneously recalled by the children after the ride.}
\end{figure}

\textbf{Recollection of contextual information.} 
We first compared the number of locations that children recollected in the pre-study with those they recalled after the experiment, as shown in Figure~\ref{fig:number_POI_child}. Since the route the families took were frequent routes to the children, it ensures a fair basis for comparison between these two measures. A paired-samples t-test confirmed that the number of locations recalled post-study (M = 3.9, SD = 1.3) was significantly higher than the pre-study baseline (M = 0.9, SD = 0.74), t(9) = -6.068, $p < .001$ (see Figure~\ref{fig:number_POI_child}).

We acknowledge the methodological limitation that this compares a measure rooted in long-term memory with one of short-term memory. However, we interpret this significant increase as evidence that the SCENIC intervention successfully drew children's attention to and created memorable experiences around a broader set of environmental features than those already prominent in their minds. A breakdown of the post-study recall shows that, on average, 2.9 (SD = 0.99) of the recalled locations were prompted by SCENIC, while 1.0 (SD = 0.67) were noticed independently by the children.

\begin{figure}[htp]
    \centering
    \includegraphics[width=0.95\linewidth]{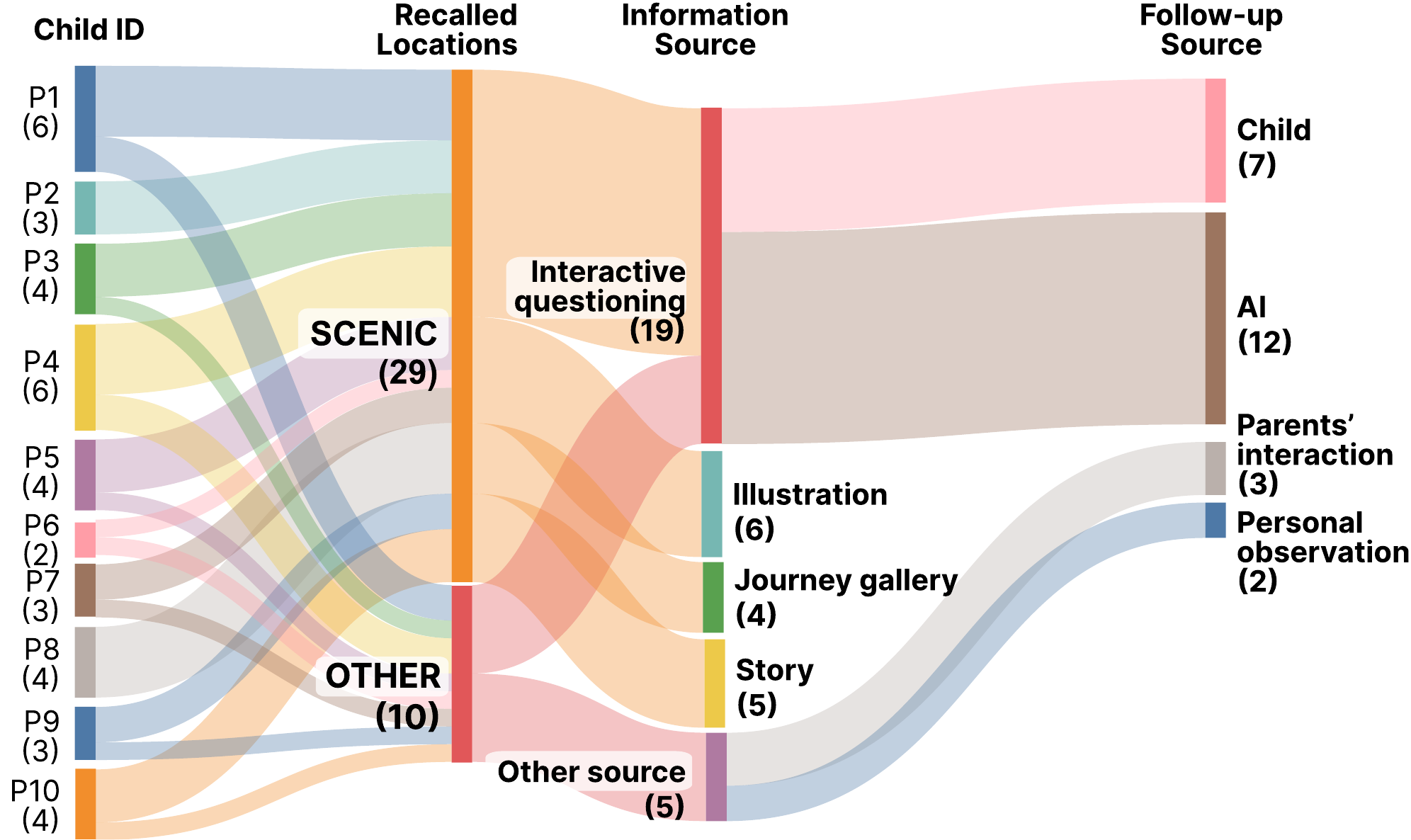}
    \caption{Sankey diagram illustrating the flow from the initial source of recalled locations (either prompted by SCENIC or independently observed) to the associated information source (e.g., interactive questioning, illustration, storytelling, journey gallery, or other sources). It then shows the follow-up source, which identifies the origin of the subsequent interaction or inquiry (e.g., Child, SCENIC, parent’s interaction, or personal observation). The numbers in parentheses represent the frequency of each category.}
    \label{fig:sankey}
    \vspace{-0.1in}
    \Description{Figure 8 shows a Sankey diagram visualizing the flow of recalled locations across three stages. The first stage identifies the origin of each recalled location, categorized as either SCENIC-prompted (29 instances) or independently recalled by children (OTHER, 10 instances). The second stage maps each recalled location to the type of information source that supported recall, including interactive questioning (19), illustration (6), storytelling (5), journey gallery (4), and other sources (5). The third stage shows the follow-up source, indicating who initiated the subsequent reflection or inquiry: the AI system (12), the child (7), parents’ interaction (3), or personal observation (2).}
\end{figure}

To further understand the pathways leading to location recall and subsequent engagement, we analyzed the flow from the initial origin of recalled locations (i.e., whether prompted by SCENIC or independently observed), through the associated information source (e.g., specific SCENIC features such as interactive questioning, illustration, storytelling, journey gallery, or other sources), and to the follow-up source, indicating who or what initiated the subsequent related interaction or inquiry (e.g., Child, SCENIC, parent’s interaction, or personal observation). This flow is visualized in the Sankey diagram presented in Figure~\ref{fig:sankey}. The numbers annotated in the figure represent the frequency of each category, with, for example, P1 (6) indicating the number of locations recalled by P1, and SCENIC (29) representing the number of locations sourced from SCENIC.

The results first demonstrate that SCENIC was the primary driver of location recall, with interactions prompted by the system accounting for the majority of remembered locations (29 out of 39, or 74.4\%). Within these SCENIC-prompted recollections, the interactive questioning feature was a significant contributor, serving as the information source in 48.3\% of these cases (14 out of 29). This highlights the importance of active, dialogic engagement for memory consolidation within the SCENIC framework.

Crucially, the diagram indicates that SCENIC fosters broader environmental awareness rather than confining attention. Over a quarter of recalled locations (10 out of 39, or 25.6\%) were those children noticed independently. Even more significantly, for half of these independently-observed recalls (5 out of 10), children chose to use SCENIC's interactive questioning feature to learn more about them (e.g., \textit{C10: "Why is this road called that?"}). This demonstrates a key success: children learned to use SCENIC not just as a content-delivery system, but as a general-purpose inquiry tool for their own curiosity. This capacity for fostering spontaneous inquiry was also seen within SCENIC-prompted interactions, with children taking the initiative to ask follow-up questions (e.g., \textit{C8: "What kind of bird is a migratory bird?"} when passing a wetland.).

Finally, the "Follow-up Source" column reveals that SCENIC successfully integrated into the family dynamics without monopolizing it, creating a healthy, hybrid ecosystem of interaction. While system-initiated interactions ("AI", 12 instances) were prominent, it is noteworthy that child-initiated inquiries ("Child", 7 instances) were also a major component, suggesting the system empowers children's conversational agency. Importantly, SCENIC did not replace other natural interactions---parental interaction and personal observation still contributed to the interaction process.

\subsubsection{RQ3: How does SCENIC System Influence Children’s Interactions with the External Environment?}
\mbox{}\\
To understand how SCENIC influences children's interactions with their surroundings, we analyzed the qualitative data from post-drive interviews, linking children's recall and experiences to the specific system features that enabled them. Table~\ref{tab:manifestations_text} provides a summary of how each core feature of SCENIC shaped the interaction with the external environment. In this section, we analyze the performance of the six cognitive development strategies proposed by SCENIC, focusing on how these strategies support children’s cognitive growth during their interactions with the system. This is particularly important because the strategies represent the design feature with the primary interaction function between children and the SCENIC system. Furthermore, as revealed in Figure~\ref{fig:sankey}, the results indicate that this feature provides the richest variety of information sources that children can recall after the journey.

\begin{table*}[ht]
\centering
\caption{How does SCENIC system provides supportive information in children’s perception of POIs.}
\label{tab:manifestations_text}
\begin{tabular}
{p{0.2\textwidth}p{0.35\textwidth}p{0.35\textwidth}} 
\toprule
\textbf{Feature} & \textbf{Empirical evidence} & \textbf{Example (from post-experiment interview)}\\ 
\midrule
Interactive questioning (initiated by AI) & Prompts children to think about their surroundings through situational questions. & "I remember the university we passed by because AI asked me what I would like to study there."\\
\midrule
Interactive questioning (initiated by the child) & Answers children’s questions about locations or interesting landmarks along the way. & "I remember the Flower Harbor Fish Viewing because I asked AI, 'What kind of fish is in the Flower Harbor Fish Viewing?'"\\
\midrule
Story & Introduces a place through a storytelling format. & "I remember 'Luting Bridge' because the AI told me the origin of its name."\\
\midrule
\multirow{2}{0.2\textwidth}{Illustration} & Includes characters chosen by children. & "I remember the library because I saw the rabbit I chose reading a book."\\
\cline{2-3}
 & Assists children in answering questions. & "I’m not sure what kind of ice cream it would create, but the generated image showed pinecone-flavored ice cream, so I answered that way."\\
\midrule
Travel memory & Features a character chosen by the child in typical location settings. & "The cat is walking on Su Causeway, it was so cool."\\
\bottomrule
\end{tabular}
\end{table*}

Specifically, we encoded the interaction logs according to the following criteria: “number of questions posed by the system – number of questions answered by children – number of answers related to images – number of questions recalled during the post-experiment interview,” with detailed results listed in Supplementary material F. For the questions that were not answered, children typically either asked SCENIC about topics that triggered their interest or engaged in conversations with their parents about the passing POIs. This aligns with our aim of treating the SCENIC strategies as scaffolds to guide children’s thinking about the POIs, rather than restricting them to only the system’s prompts, as doing so may lead to negative feelings like test or fatigue.

Among the answered questions, those related to the provided images were most commonly associated with creativity (i.e., scenario-based role-playing and expanded thinking). In contrast, when answering logic or decision-making questions, children typically relied on their own reasoning abilities or simply stated “I don’t know”.

Although only 15.3\% of the generated questions were mentioned in the post-experiment interviews, the answers covered both factual recall and productive recall, aligning with the goals of informal learning \cite{cronin2024factors}. For example, \textit{``I remember the origin of the Su Causeway’s name."} (C5) reflects factual recall, while \textit{``I remember the art museum because my answer was that musical elements would give paintings vivid colors."} (C2) is an example of productive recall.
\section{Discussion}
\subsection{Location-based Learning Systems for Children}



This study contributes to the field of contextual learning by extending support systems to the challenging, yet potentially rich, environment of car rides via SCENIC. Unlike more structured contextual learning settings such as museums \cite{cronin2024factors} or parks \cite{zhang2023observe}, the in-car environment is characterized by fragmented exposure to external stimuli, rapid visual changes, and limited user control over pacing or focus. These inherent constraints demand specific design considerations distinct from those in more static learning environments.

Beyond the specific application in private vehicles, the principles underlying SCENIC may offer insights for other mobile, or contextual learning contexts. The challenge of engaging learners with fragmented, dynamic stimuli is also present in public transit scenarios (buses, trains) or even during guided tours in complex environments. The core idea of using generative AI to weave disparate points of information into a coherent narrative or conversational thread, coupled with strategies that scaffold thinking from multiple cognitive angles (as exemplified by the SCENIC prompts), could be broadly applicable. Future work could explore adapting these strategies to different mobile contexts, potentially integrating sensor data beyond location (e.g., speed, weather) for richer contextualization. Furthermore, investigating how these AI-driven scaffolding techniques influence not just recall and engagement, but also the transferability of the thinking skills (e.g., classification, inference, perspective-taking) fostered during the journey to other real-world situations.

\subsection{Children's Experience during Car Rides}
Children's boredom during car rides is a well-documented issue in car rides \cite{hoffman2013car, price2013travel, inbar2011make}. However, in this study, we trying to enhance the engagement through personalization, freedom of choice and immersion by not relying on prolonged screen time as a solution. We take various measures to enhance children’s experience before, during, and after the car ride. Before the journey, children are allowed to choose a character to represent them, which encourages their active participation and engagement with the system, while also introducing content variety to routes that may be repetitive. Additionally, the continuity of the selected character throughout the journey helps strengthen the children’s sense of identity and trust in the system.

To address the challenges children face during car rides, such as limited physical space and the fleeting nature of the changing scenery, we implement visual information compensation by illustrating the character within the scene. This helps children imagine and relate to the scenes within the POIs.

To enhance children’s recollection of the journey after it concludes, we generate journey memory images, assisting children in forming visual memories of their experiences.

Furthermore, beyond the SCENIC system’s core design features, we encourage spontaneous creativity and exploration by allowing children to ask and answer questions not only about the POIs selected by SCENIC but also about other points of interest they encounter during the ride. This open-ended interaction fosters curiosity and empowers children to take an active role in the experience.

\subsection{SCENIC's Influence on Family Time during Car Rides}

In our experiment, we observed that parents employed three primary intervention strategies to guide their children in using the SCENIC system. First, parents assisted children in developing an accurate understanding of both the capabilities and limitations of AI, encouraging them to critically interpret AI responses and to clearly define the role of AI in the interaction. Second, parents actively encouraged children to ask questions, gradually fostering their ability to interact naturally and fluently with the system. Third, when children encountered difficulties in responding to AI prompts, parents provided timely assistance, thereby ensuring a smooth and continuous communication process.

Moreover, the three families with multiple children participating in the experiment demonstrated harmonious and fluid usage of the SCENIC system. The interaction within these families generally involved either alternating turns between two children in responding to questions (UF3, UF5) or one child responding while the other primarily listened and provided feedback (UF6).

In summary, the SCENIC system not only preserved the valuable family time during car rides, but also served as a mediator, introducing a novel topic of conversation during family journeys.

\section{Limitations and Future Work}
\textbf{Long-term experimentation.}
Our study provides valuable initial insights into the potential of generative AI for in-car cognitive development. But it is important to acknowledge the primary limitations---its short-term, single-session evaluation design. While our field study demonstrate immediate positive effects on engagement and location recall, this methodology cannot determine the long-term impact of SCENIC on children's cognitive development. Key unanswered questions include whether the observed engagement is sustained over repeated use or is primarily a result of the system's novelty, and whether the  cognitive skills prompted by SCENIC (e.g. inference, perspective-taking) transfer to other contexts over time.

To address these questions, future work should involve a longitudinal study. For instance, a multi-week deployment with families. Such a study would also need to account for confounding variables like home and school education and individual differences among children to better isolate the system's long-term effects.

\textbf{Robustness of the system in different regions.}
Considering the convenience of recruiting participants, our current experiment utilized eight driving routes primarily conducted in a single city in southern China. In the future, we plan to conduct field experiments across diverse environments, including areas with varying POI densities as well as both urban and rural settings, to further validate the applicability and robustness of the SCENIC system under different regional conditions.

\textbf{Integrating with more devices.}
In the current design, we opted for a web-based interface accessed via iPad tablets to simplify implementation and facilitate early-stage user engagement. During the user experiments, several families (UF4, UF6, UF7) expressed interest in having the system deployed on children’s smartwatches. We believe that migrating the core functionalities of the SCENIC system—such as the initial personalized design selection, the interactive conversation with supporting illustrations during the process, and the concluding travel recollections——onto a wearable device would not only preserve its primary strengths but also further lower the barrier to use. In the future, we will explore the technical details and permission requirements specific to this direction.

\section{Conclusion}
In this paper, we presented SCENIC, a location-based system designed to foster cognitive development for children during car rides. Recognizing the unique challenges of this context——transient stimuli and limited parental capacity for interaction due to driving demands and knowledge gaps——we developed tailored strategies. SCENIC employs cognitive prompting strategies anchored to points of interest, aiming to scaffold children's creativity, logical reasoning, and decision-making about their surroundings. These are integrated with location-based storytelling and a journey gallery to enhance engagement. Evaluations of the generated-content quality demonstrated that SCENIC's generated prompts and language appropriateness significantly outperformed baseline methods and parent-authored examples in key dimensions like contextual relevance and cognitive significance. Furthermore, the user study indicated SCENIC's positive impact on increasing children's recall of route locations and fostering awareness of their environment. This work highlights the potential of context-aware generative systems to transform passive travel time into meaningful opportunities for cognitive engagement. 

\begin{acks}
We thank all participating families for their time and the anonymous reviewers for their valuable feedback. This research was supported by the National Natural Science Foundation of China: 62207023.
\end{acks}

\bibliographystyle{ACM-Reference-Format}
\bibliography{acmart}

\end{document}